%% file: Perra.Pak.AE26.tex
\documentclass[final,3p,times,twocolumn]{elsarticle}

\usepackage{graphicx}      
\usepackage{natbib}        
\usepackage{amsmath}
\usepackage{svg}
\usepackage{subcaption}
\usepackage{amsthm}
\usepackage{amssymb,mathtools}
\usepackage{float}
\usepackage{siunitx}
\usepackage{bm}
\usepackage{xurl}
\usepackage{needspace}
\usepackage{thmtools}
\usepackage{placeins}
\usepackage{amsthm}

\newtheorem{prob}{Problem}

\makeatletter
\def\@endtheorem{\par}
\makeatother

\usepackage{eso-pic}
\AddToShipoutPictureBG*{%
	\AtPageUpperLeft{%
		\setlength\unitlength{1in}%
		\hspace*{\dimexpr0.5\paperwidth\relax}
		\makebox(0,-1.75)[c]{
			\begin{tabular}{c c}
				Uto Perra \emph{et al.}, \\
				Optimization Models and Steady-State Minimum-Fuel Operating Strategies for Hydrogen-based Hybrid Electric Aerospace Propulsion Systems, \\
				uploaded to ArXiv \today \\
			\end{tabular}}}}

\journal{}
\makeatletter
\def\ps@pprintTitle{%
	\let\@oddhead\@empty
	\let\@evenhead\@empty
	\let\@oddfoot\@empty
	\let\@evenfoot\@empty
}
\makeatother

\begin{document}

\begin{frontmatter}
	
\title{Optimization Models and Steady-State Minimum-Fuel Operating Strategies for Hydrogen-based Hybrid Electric Aerospace Propulsion Systems}


\author{Uto~Perra\corref{correspondingAuthor}\fnref{TUe}}
\ead{u.perra@tue.nl}
\cortext[correspondingAuthor]{Corresponding author.}
\author{Faezeh~Pak\fnref{TUe}}
\author{Evangelia~Pontika\fnref{Cranfield}}
\author{Sahil~Bhapkar\fnref{DLR-EL}}
\author{Daniel~Ewald\fnref{KIT}}
\author{Dario~Buzzola\fnref{UniGe}}
\author{Theo~Hofman\fnref{TUe}}
\author{Frank~Willems\fnref{TUe}}
\author{Mauro~Salazar\fnref{TUe}}
\affiliation[TUe]{organization={Department of Mechanical Engineering, Eindhoven University of Technology},
	addressline={Groene Loper 3},
	postcode={5612 AE}, 
	city={Eindhoven}, 
	country={The Netherlands}
}

\affiliation[Cranfield]{organization={Cranfield University},
	addressline={College Road},
	postcode={MK43 0AL}, 
	city={Cranfield}, 
	country={United Kingdom}
}

\affiliation[DLR-EL]{organization={DLR-Institut für Elektrifizierte Luftfahrtantriebe},
	addressline={Lieberoser Straße 13a},
	postcode={03046}, 
	city={Cottbus}, 
	country={Germany}
}

\affiliation[KIT]{organization={Karlsruhe Institute of Technology},
	addressline={Adenauerring 20b},
	postcode={76131}, 
	city={Karlsruhe}, 
	country={Germany}
}

\affiliation[UniGe]{organization={Università degli Studi di Genova},
	addressline={Via Montallegro, 1},
	postcode={16145}, 
	city={Genova}, 
	country={Italy}
}



\begin{abstract}
\noindent This paper presents an optimization framework for the operation of hydrogen-based hybrid electric aerospace propulsion systems consisting of a hydrogen gas turbine and an electric motor powered by a solid oxide fuel cell, which is in turn connected to the gas turbine via multiple gas channels and heat exchangers.
Our framework computes the minimum-fuel optimal operating strategies over a flight mission accounting for the complex propulsion system with strong thermodynamic and mechanical coupling between components.
Specifically, we first identify surrogate optimization models of the components employing high-fidelity model simulations.
Second, we frame the minimum-fuel optimal control problem over a given flight mission and parse it into a static nonlinear optimization problem that can be efficiently solved with off-the-shelf nonlinear programming algorithms.
Finally, we showcase the potential of our optimization framework by applying it to a typical flight mission of an advanced, commuter aircraft (Beechcraft 1900D market segment), considering a parallel propulsion system architecture with four different configurations that share a common baseline but differ in the inclusion of an additional battery and by-pass valves around the two heat exchangers. 
We validate the resulting optimal trajectories against high-fidelity simulation results, demonstrating the accuracy of our framework.
Our results on the use case show that adding by-pass valves around the air and hydrogen heat exchangers can reduce fuel consumption by 19.11~$\%$ for the case without the battery, and by 19.56$\%$ when the battery is integrated.
At the same time, we show that adding a battery yields a slight increase in fuel consumption (below 1\%) for future projected energy densities under steady-state operating conditions. Conversely, when considering state-of-the art energy densities, the additional battery weight outweighs the benefits, limiting its potential applicability to only assisting transients, which are not considered in the present work.
\end{abstract}


\begin{highlights}
\item This paper presents optimization models and minimum-fuel optimal operating strategies for hydrogen-based hybrid electric aero engines under steady-state conditions.

\item Four different propulsion system configurations that share a common baseline are analyzed and compared in terms of operating strategies and total fuel consumption over a given flight mission.

\item The resulting optimal trajectories are validated against high-fidelity model simulations, showing high model accuracy in the most frequent operating windows, while achieving the optimal solution in approximately 3 second.

\item A case study on a typical commuter aircaft flight mission shows the benefit of using heat exchanger by-pass valves, yielding a fuel consumption reduction by more than 19~$\%$, whereas the inclusion of a battery provides no additional fuel savings due to the associated mass penalty.
\end{highlights}


\end{frontmatter}

\input{Sections/introduction.tex}

\input{Sections/methodology.tex}
\input{Sections/results.tex}

\input{Sections/conclusions.tex}

\section*{Acknowledgment}
\noindent
We thank Dr.\ I.\ New and Dr.\ O.\ Borsboom for proofreading this paper.

\section*{Declaration of generative AI}
\noindent
The authors used ChatGPT to improve the clarity and structure of selected parts of the manuscript, including the abstract, conclusions, and optimization results sections.
\section*{Author Contributions}
\noindent
\textbf{Uto Perra:}~Conceptualization, Data Curation, Formal Analysis, Investigation, Methodology, Software, Writing -- original draft, Writing -- review and editing.
\textbf{Faezeh Pak:}~Conceptualization, Data Curation, Investigation, Software, Writing -- original draft, Writing -- review and editing.
\textbf{Evangelia Pontika:}~Data Curation, Formal Analysis, Investigation, Validation, Writing -- review and editing.
\textbf{Sahil Bhapkar:}~Data Curation, Investigation, Validation, Writing -- review and editing.
\textbf{Daniel Ewald:}~Data Curation, Software, Writing -- review and editing.
\textbf{Dario Buzzola:}~Data Curation, Software, Writing -- review and editing.
\textbf{Theo Hofman:}~Writing -- review and editing, Supervision.
\textbf{Frank Willems:}~Writing -- review and editing, Supervision.
\textbf{Mauro Salazar:}~Writing -- review and editing, Funding acquisition, Supervision.

\section*{Funding}
\noindent
Funded by the European Union under grant number 101138488 and by the UK Research and Innovation (UKRI) funding guarantee under the project reference 10106893. Views and opinions expressed are however those of the author(s) only and do not necessarily reflect those of the European Union. Neither the European Union nor the granting authority can be held responsible for them.

\bibliographystyle{elsarticle-num} 
\bibliography{Bibliography/AE26.bib}




\end{document}

%% file: Sections/introduction.tex
\section{Introduction}
\begin{figure*}[!t]
	\centering
	\includegraphics[width=\textwidth]{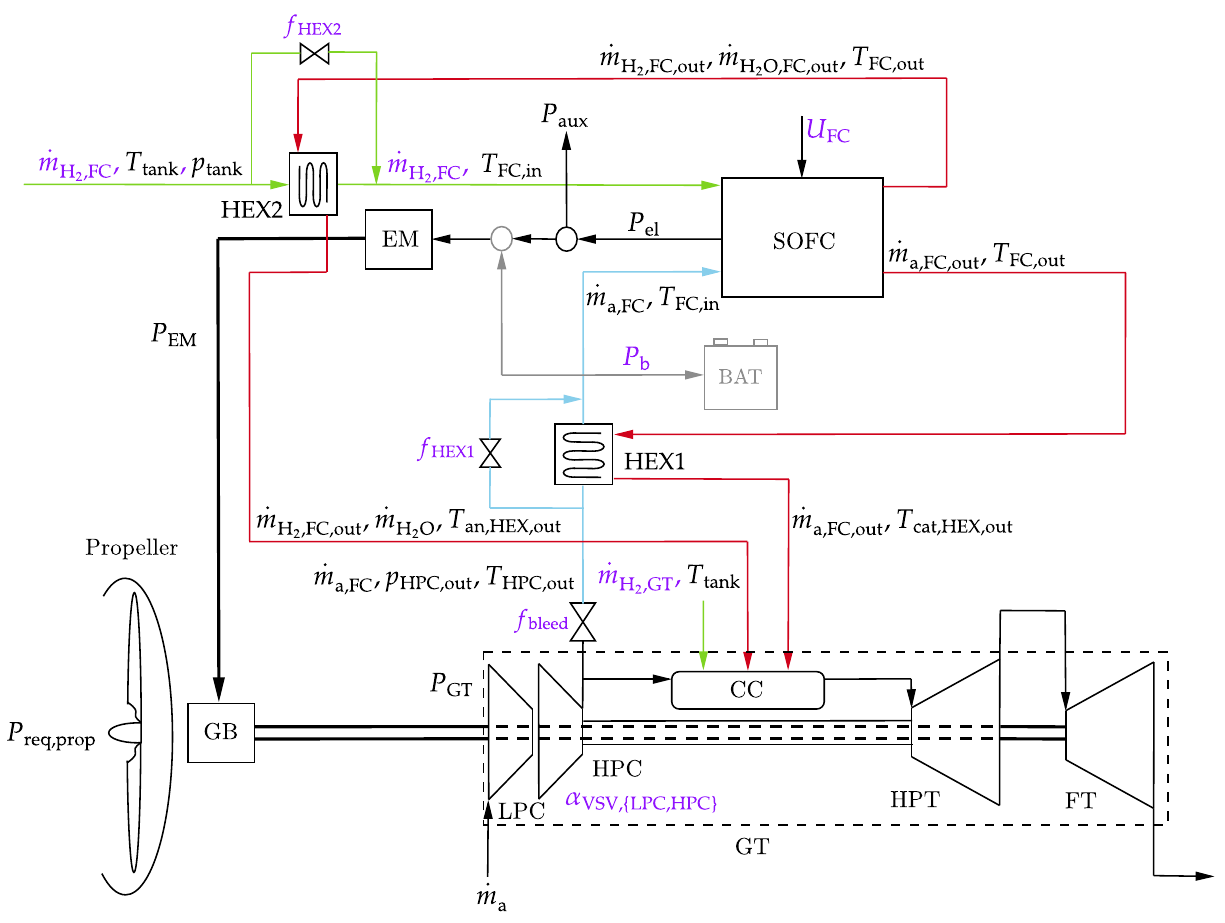}
	\caption{Simplified scheme of the examined hybrid propulsion system. The solid oxide fuel cell (SOFC) stack is tightly integrated with a gas turbine (GT). A bleed valve with opening $f_\mathrm{bleed}$ regulates the air flow at the outlet of the high-pressure compressor (HPC) directed to the SOFC (light blue line), which is preheated via a heat exchanger (HEX1). The hot off-gas from the SOFC (red lines) from each side then undergoes heat transfer through the heat exchangers before being sent to the combustion chamber (CC). Two by-pass valves, each per heat exchanger, with openings $f_\mathrm{HEX1}$ and $f_\mathrm{HEX2}$, are included to the baseline architecture improve the temperature control. The hot gases then expand in the high-pressure turbine (HPT), which drives the high-pressure compressor and the low-pressure compressor (LPC). Variable stator vanes for both the low pressure compressor $\alpha_\mathrm{VSV,LPC}$ and high pressure compressor $\alpha_\mathrm{VSV,HPC}$ are included to prevent instabilities, extending the stable operating envelope. Finally, the gases expand in the free turbine (FT), which then drives the propeller. The purple variables correspond to the control variables, while the green lines correspond to the hydrogen flows. The grey lines correspond to the configuration integrating a battery, where the battery power $P_\mathrm{b}$ denotes the power supplied to or extracted from the battery and serves as an additional control variable.}
	\label{fig:IPPS-scheme}
\end{figure*}

The continuous growth of the aerospace sector, together with increasing travel demand, is leading to a significant rise in carbon-related emissions, which are expected to grow even further in the coming years. As reported in \cite{easa2025}, aviation accounted for approximately 2.5~$\%$ of global CO$_2$ emissions in 2023, and 47~$\%$ of the toal aviation CO$_2$ emissions between 1940 and 2019 occurred since 2000. To mitigate this issue, researchers have investigated different solutions. One promising approach to eliminate carbon-related operational emissions while having a high efficiency is using   hydrogen fuel in combination with tightly-coupled hybrid systems composed of a gas turbine (GT) engine and solid oxide fuel cell (SOFC) stacks where the SOFC both becomes part of the gas turbine cycle (thermodynamic coupling) and provides electrical power (mechanical coupling), as shown by \cite{SanBenito2025} and \cite{MashambaEtAl2024}. However, to fully exploit the potential of such a hybrid configuration and to balance the penalty of increased weight, the system must be operated as efficiently as possible.  Therefore, the hybrid system requires fuel-optimal operating strategies, which in turn necessitate accurate models. Nevertheless, the complexity of the system and the tight coupling between its components affect the operation of the propulsion system and represent a significant challenge.
To this end, we aim to frame a constrained minimum-fuel optimization problem, subject to components behavior, operational requirements, and safety constraints, while capturing the strong thermodynamic and mechanical coupling between the different subsystems. To accurately represent the behavior of the hybrid system, high-fidelity SOFC models \cite{AnderssonEtAl2012,NajafiEtAl2024} or the use of gas turbine components maps \cite{TsoutsanisEtAl2014,AlbertoEtAl2014} are often employed. 
However, solving the optimization problem using computationally expensive, often iterative, high-fidelity models makes it intractable. Moreover, components maps for gas turbines are typically proprietary and confidential, which limits their direct use. A potential approach is the use of accurate yet tractable reduced-order surrogate models. In this paper, we present steady-state minimum-fuel operating strategies for hydrogen-based hybrid electric aero engines, employing reduced-order surrogate models of the individual components, capturing the thermodynamic couplings between them, providing a balance between accuracy and computational cost, and we compare the performance to that of the same propulsion system integrated with a battery. \\
\indent \emph{Related literature}: 
This paper relates to two main research streams: 1) energy management of hybrid propulsion systems, and 2) modeling and optimization of hybrid GT/SOFC systems. In the first stream, the authors of \cite{BrahmaEtAl2000}, \cite{SundstroemGuzzellaEtAl2008}, and \cite{RobuschiEtAl2020} presented models and algorithms to compute minimum-fuel energy management strategies for ground hybrid electric powertrains over given driving cycles. Another modeling approach was proposed in \cite{TurkerEtAl2022}, which investigated the energy management strategy using a neural-network-based surrogate model for
a range-extended vehicle. Although different modeling and optimization approaches have been explored, these studies focus on ground vehicles and are not directly applicable to either aerospace propulsion systems or hybrid systems combining gas turbine and fuel cells. However, the growing interest in sustainable aviation has motivated the application of optimal energy management strategies to the aerospace domain. In this context, \cite{DoffEtAl2020}, \cite{XieEtAl2022}, \cite{ZhangEtAl2022}, and \cite{MisleyEtAl2021}  presented optimal energy management strategies for hybrid-electric aircraft employing convex optimization, model predictive control, and the Pontryagin minimum principle. Similarly, \cite{FerrulliEtAl2024} investigated the optimal energy management for a fuel-cell hybrid-electric aircraft based on dynamic programming. In \cite{GraziosoEtAl2025} the authors investigated the mission-level optimization of tri-source hybrid electric aircraft, combining a gas-turbine engine, hydrogen fuel-cell systems, and batteries enabling power allocation in each flight segment over a given flight mission. Their results showed a 31.5 $\%$ total fuel reduction for a 300 nmi mission compared to the reference aircraft. However, despite the integration of batteries, which play a relevant role during transient operations, these studies focus on hybrid-electric propulsion systems that do not include tightly thermodynamically coupled components, such as gas turbines and solid oxide fuel cells, which are the core of the present work.

The second stream addresses the modeling and optimization of hybrid GT/SOFC systems. In this stream, \cite{SafariEtAl2020} performed a multi-objective optimization of the overall output power and system efficiency of an SOFC/GT heat and power system, comparing the performance of two heuristic algorithms: genetic algorithm and particle swarm optimization. In \cite{FredrikssonEtAl2004} the optimal combination of key decision variables is  investigated, including air and fuel flow rates, cell voltage, and air inlet temperature, to maximize the design-point efficiency of a GT/SOFC system using a genetic algorithm. In \cite{FinocchiEtAl2022} the authors proposed a surrogate-based robust optimal design approach for a hybrid GT/SOFC system aiming to achieve high efficiency with low performance variability under fuel composition variations. However, these studies focus on finding the optimal design of the hybrid system and do not study the optimal operation over a flight mission.
In \cite{ChenEtAl2024} the authors studied the design optimization of a hybrid GT/SOFC system, focusing on key design parameters, and applied to a long-endurance unmanned aerial vehicles. To manage the power demand across different flight phases, they adopted a fuzzy-logic-based energy management strategy. In contrast, the present work employs an optimization-based approach to derive the minimum-fuel optimal strategies. Kameswaran et al. \cite{KameswaranEtAl2007} proposed a dynamic optimization framework to determine optimal control policies for hybrid fuel cell-gas turbine power systems. Although their approach employs detailed dynamic models, it does not consider the recirculation of gases from the SOFC to the GT, which is a crucial part of the present work. Moreover, their study does not compare the optimal control strategies with those obtained for the baseline system including a battery or heat exchangers by-pass valves, where the latter are crucial for temperature control. 

In conclusion, to the best of the authors' knowledge, no previous studies have investigated and compared the steady-state minimum-fuel operation of hydrogen-based hybrid electric GT-SOFC aerospace propulsion systems with that of the same system integrating: (i) a battery
	only, (ii) heat exchangers by-pass valves only, and
	(iii) both a battery and heat exchangers by-pass
	valves, while capturing the strong thermodynamic couplings
	among components through surrogate models.

\begin{figure}
    \centering
    \includegraphics[width=\columnwidth]{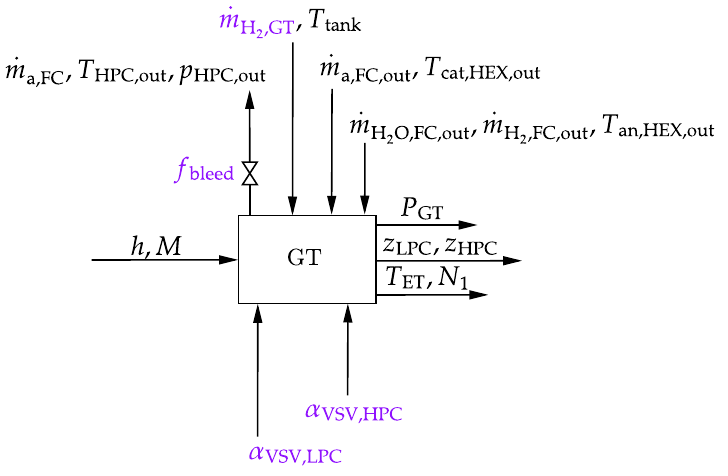}
    \caption{Simplified scheme representing the GT interfaces as a function of the operating flight altitude $h$ and Mach number $M$. A fraction of the total mass flow rate leaving the high-pressure compressor, determined by the bleed fraction $f_\mathrm{bleed}$, is directed to the SOFC. This portion of the mass flow is referred to as $\dot{m}_\mathrm{a,FC}$, and leaves the HPC at temperature $T_\mathrm{HPC,out}$ and pressure $p_\mathrm{HPC,out}$. In return, the GT receives from the SOFC the outlet air flow $\dot{m}_\mathrm{a,FC,out}$ at temperature $T_\mathrm{cat,HEX,out}$, together with the unutilized hydrogen $\dot{m}_\mathrm{H_2,FC,out}$ and the steam $\dot{m}_\mathrm{H_2O}$ at a temperature $T_\mathrm{an,HEX,out}$. Additionally, an extra hydrogen flow $\dot{m}_\mathrm{H_2,GT}$ is injected from the storage tank at temperature $T_\mathrm{tank}$. As a result, the GT is providing a mechanical power $P_\mathrm{GT}$.} 
    \label{fig:GT interfaces}
\end{figure}
\emph{Statement of contributions}: In order to address this research gap, this paper presents component surrogate models and minimum-fuel optimization strategies for hydrogen-based hybrid-electric aerospace propulsion systems for the four different configurations, capturing the tight thermodynamic couplings between the components. First, we carry out a design of experiments and we identify surrogate models of the gas turbine and SOFC, along with models of the heat exchangers and the battery pack. Second, we frame a minimum-fuel optimization problem for the four different propulsion systems by leveraging and coupling the identified models, together with safety and operational constraints at component level and system integration level. Third, we solve the problems, and we validate our approach by comparing the optimization results with those obtained from the high-fidelity models that we used to derive the surrogate models. Finally, we compare the four propulsion configurations in terms of performance and total fuel consumption.

\emph{Organization}: The remainder of this paper is organized as follows: Section 2 presents the components' surrogate models and the minimum-fuel optimization problems for the two propulsion configurations, which include the identified models, operational requirements, and safety constraints. In Section 3 we present and validate the results against high-fidelity models, demonstrating the accuracy of our approach. Moreover, we compare the total fuel consumption of the four propulsion configurations, highlighting the impact of using a battery and by-pass valves. Finally, in Section 4, we draw the conclusions and provide an outlook for future research. 

%% file: Sections/methodology.tex
\section{Methodology}
In this section, we first identify the individual component surrogate models of the hybrid propulsion system shown in Fig.~\ref{fig:IPPS-scheme}, as function of the control inputs, denoted by the purple variables, and the interfaces with the rest of the system. The surrogate models are constructed using a quadratic structure, which strikes a balance between accuracy and computational tractability, while capturing the main nonlinear interactions among variables and remaining compatible with efficient nonlinear programming solvers. Second, we construct the minimum-fuel optimization problems for the four configurations employing these models together with safety and operational constraints.

The considered hybrid propulsion system is composed of a gas turbine, a solid oxide fuel cell stack, two heat exchangers that preheat the SOFC inlet air and fuel flows  (HEX1 and HEX2), and an electric system with converter, inverter, and motor (EM). The SOFC is tightly coupled to the GT, which provides hot and pressurized air. This air flow is controlled by a bleed valve with opening $f_\mathrm{bleed}$ and further heated in a heat exchanger HEX1 that recovers thermal energy from the SOFC cathode exhaust gases.  Similarly, the fuel flow entering the SOFC is heated in a second heat exchanger HEX2 using the thermal energy contained in the SOFC anode exhaust gases. To allow better temperature control, by-pass valves with openings $f_\mathrm{HEX1}$ and $f_\mathrm{HEX2}$ are included alongside the air and hydrogen heat exchangers, respectively. To enhance overall system efficiency, the anode and cathode exhaust gases  are directed to the combustion chamber of the GT to exploit the unutilized hydrogen and the enthalpy of the gases, after passing through  the heat exchangers.

\subsection{Objective}
\begin{figure}[t]
	\centering
	\includegraphics[width=\columnwidth]{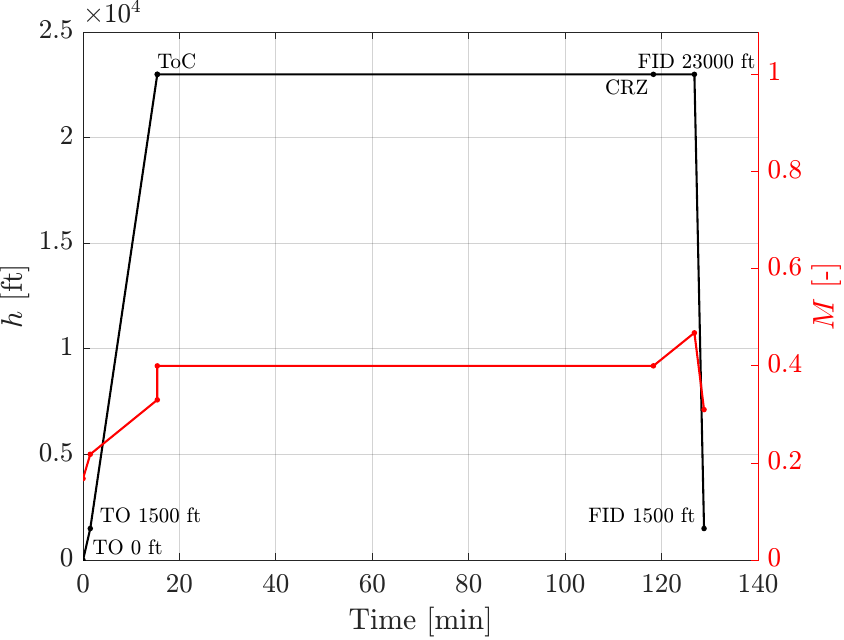}
	\caption{Simplified scheme representing the flight altitude $h$ and Mach number $M$ as a function of time. The dots denote the operating points considered: Take-off at 0 ft (TO 0 ft) and 1500 ft (TO 1500 ft), top of climb (ToC), cruise (CRZ), flight idle at 23000 ft (FID 23000 ft) and at 1500 ft (FID 1500 ft).} 
	\label{fig:Mission}
\end{figure}
To construct the minimum-fuel optimization problem, we first define the objective of our problem. The main objective in this problem is the minimization of the total hydrogen consumed by the propulsion system over a set of given flight operating conditions (Fig.~\ref{fig:Mission}), which can be expressed as: 
\begin{equation}
	\min \int_{0}^{t_\mathrm{fin}}{\dot{m}_{\mathrm{H_2,tot}}}(t) \mathrm{d} t, 
	\label{eq: objective}
\end{equation}
where $\dot{m}_{\mathrm{H_2,tot}}$ is equal to the sum of the fuel injected in the SOFC and the fuel injected within the GT
\begin{equation}
	\dot{m}_{\mathrm{H_2,tot}}(t) = \dot{m}_\mathrm{H_2,FC}(t) + \dot{m}_\mathrm{H_2,GT}(t),
	\label{eq: total fuel}
\end{equation}
and $t_\mathrm{fin}$ is the final time instant. To simplify the notation, the time dependence is omitted hereafter. 
\subsection{Gas Turbine}
In this section, we present the gas turbine models and outline the procedure used to derive them. 
The goal of the GT surrogate models is to analytically but still accurately represent the behavior of the gas turbine as function of the power delivered, its control variables, and its interfaces with the rest of the propulsion system, as shown in Fig.~\ref{fig:GT interfaces}. 

\emph{Design of Experiments and Surrogate Modeling Formulation}:
To construct the surrogate models, we first perform high-fidelity gas turbine simulations within the identification space using Turbomatch\footnote[1]{Turbomatch is a component-based simulation tool which performs thermodynamic calculations and resolves the enthalpy balance, mass balance and power balance using well-established gas turbine performance methods available in literature and textbooks \cite{Walsh2004,Kurzke2018} that do not need to be iterated in this paper.}, an in-house gas turbine performance simulation code developed at Cranfield University \cite{Nikolaidis2020,Janikovic2010}, following a predefined Latin Hypercube sampling plan that includes variation of the power requirements, gas turbine control variables and interfaces with the SOFC at the operating conditions given by the mission requirements (Table~\ref{tab:Mission_Parameters}).First, we fix the design point of the gas turbine at take-off with Turbine Entry Temperature ($T_\mathrm{ET}$) at \SI{1300}{\kelvin}, Overall Pressure Ratio (OPR) at 10, and Nozzle Pressure Ratio at 1.14, and the interfaces with the SOFC correspond to the design point of the SOFC with cell voltage $U_\mathrm{FC}$ = \SI{0.6}{\volt}, fuel utilization $f_\mathrm{FU}$ = 0.2, air utilization $f_\mathrm{AU}$ = 0.1, and a power split between the electric motor and GT at 20$~\%$. Subsequently, we perform the off-design performance simulations with variation of input parameters according to the Latin Hypercube input datasets and we used the results to identify the models parameters through least-squares error optimization. Specifically, we aim to minimize the mean squared error between the predicted values and the full simulation results for six operating conditions: 1. take-off at 0 ft, 2. end of take-off at 1500 ft, 3. top of climb, 4. cruise, 5. beginning of descent (flight idle) at 23000 ft, and 6. end of descent (flight idle) at 1500 ft, as shown in Fig.~\ref{fig:Mission}. The identification space is bounded by
\begin{equation}
	\dot{m}_{\mathrm{H_2,GT,tot}} \in \left[\dot{m}^\mathrm{min}_{\mathrm{H_2,GT,tot}},  \dot{m}^\mathrm{max}_{\mathrm{H_2,GT,tot}} \right],
	\label{eq:mH2gt}
\end{equation}
\begin{equation}
	\alpha_{\mathrm{VSV,LPC}} \in \left[ \alpha^\mathrm{min}_{\mathrm{VSV,LPC}}, \alpha^\mathrm{max}_{\mathrm{VSV,LPC}} \right],
	\label{eq:VSV_LPC}
\end{equation}
\begin{equation}
	\alpha_{\mathrm{VSV,HPC}} \in \left[ \alpha^\mathrm{min}_{\mathrm{VSV,HPC}}, \alpha^\mathrm{max}_{\mathrm{VSV,HPC}} \right],
	\label{eq:VSV_HPC}
\end{equation}
\begin{equation}
	\dot{m}_{\mathrm{H_2O,FC,out}} \in \left[ \dot{m}^\mathrm{min}_{\mathrm{H_2O,FC,out}},  \dot{m}^\mathrm{max}_{\mathrm{H_2O,FC,out}} \right],
	\label{eq:msteam}
\end{equation}
\begin{equation}
	T_{\mathrm{mix}} \in \left[ T^\mathrm{min}_{\mathrm{{mix}}},  T^\mathrm{max}_{\mathrm{{mix}}} \right],
	\label{eq:T_H_2,GT}
\end{equation}
\begin{equation}
	f_{\mathrm{bleed}} \in \left[ f^\mathrm{min}_{\mathrm{bleed}},  f^\mathrm{max}_{\mathrm{bleed}} \right],
	\label{eq:f_bleed}
\end{equation}
\begin{equation}
	\dot{m}_{\mathrm{a,FC,out}} \in \left[ \dot{m}^\mathrm{min}_{\mathrm{a,FC,out}},  \dot{m}^\mathrm{max}_{\mathrm{a,FC,out}} \right],
	\label{eq:ma_exit}
\end{equation}
\begin{equation}
	T_{\mathrm{cat,HEX,out}} \in \left[ T^\mathrm{min}_{\mathrm{{cat,HEX,out}}},  T^\mathrm{max}_{\mathrm{{cat,HEX,out}}} \right],
	\label{eq:T_cat_exh_HEX_out}
\end{equation}
where $\alpha_\mathrm{VSV,LPC}$ and 
$\alpha_\mathrm{VSV,HPC}$ denote the LPC and HPC variable stator vanes angle, respectively. $\dot{m}_{\mathrm{H_2,GT,tot}}$ denotes the total fuel flow entering the combustion chamber, that is, the sum of the supplied hydrogen flow $\dot{m}_{\mathrm{H_2,GT}}$ and the SOFC outlet hydrogen flow $\dot{m}_{\mathrm{H_2,FC,out}}$, therefore it is equal to
\begin{figure}[]
		\centering
		\includegraphics[width=\columnwidth]{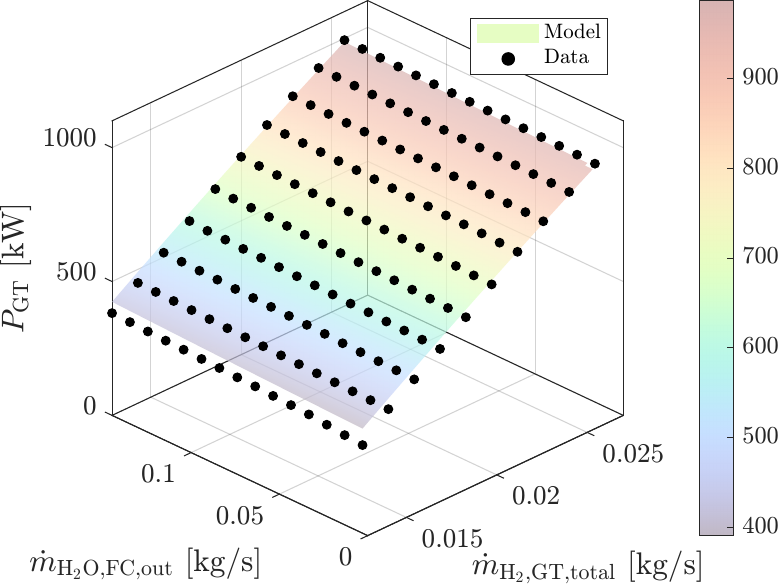}
		\caption{Comparison between the GT power $P_\mathrm{GT}$ reduced-order model and the high-fidelity results  as a function of the total hydrogen mass flow $\dot{m}_\mathrm{H_2,GT}$ and the outlet SOFC steam flow $\dot{m}_\mathrm{H_2O,FC,out}$ for constant $\alpha_\mathrm{VSV,LPC}$, $\alpha_\mathrm{VSV,HPC}$, $T_\mathrm{mix}$, $f_{\mathrm{bleed}}$,   $T_{\mathrm{cat,HEX,out}}$, and SOFC air utilization factor $f_\mathrm{AU}$.}
		\label{fig:GT_model_P_mf_fb}
\end{figure}
\begin{equation}
	 \dot{m}_{\mathrm{H_2,GT,tot}} = \dot{m}_{\mathrm{H_2,GT}} +  \dot{m}_{\mathrm{H_2,FC,out}}.
	\label{eq:mdotH2,GT,total}
\end{equation} 
$\dot{m}_\mathrm{a,FC,out}$ indicates the mass flow rate of oxygen-depleted air at the SOFC outlet, which, after undergoing heat transfer in HEX1, reaches the temperature $T_{\mathrm{cat,HEX,out}}$ and is then routed back to the gas turbine. $T_{\mathrm{mix}}$ represents the weighted average temperature of the mixture of hydrogen and steam entering the GT. This temperature results from the mixing of two streams: the hydrogen supplied from the tank, $\dot{m}_{\mathrm{H_2,GT}}$, at temperature $T_\mathrm{tank}$, and the anode off-gases exiting the SOFC, composed of hydrogen $\dot{m}_{\mathrm{H_2,FC,out}}$ and steam $\dot{m}_{\mathrm{H_2O,FC,out}}$, at temperature $T_{\mathrm{an,HEX,out}}$. Therefore, the mixture temperature is given by
\begin{equation}
	T_{\mathrm{mix}} = \frac{\sum \dot{m}_\mathrm{in} \cdot c_\mathrm{p,in}(T_\mathrm{in}) \cdot T_\mathrm{in}}{\dot{m}_\mathrm{mix} \cdot c_\mathrm{p,mix}(T_{\mathrm{mix}})},
	\label{eq:T_H_2,GT_mix}
\end{equation}
where the term $\sum \dot{m}_\mathrm{in} \cdot c_\mathrm{p,in}(T_\mathrm{in}) \cdot T_\mathrm{in}$ is equal to
\begin{equation}
	\fontsize{9.3}{12}\selectfont 
	\begin{split}
\sum \dot{m}_\mathrm{in} \cdot c_\mathrm{p,in}(T_\mathrm{in}) \cdot T_\mathrm{in} = \dot{m}_{\mathrm{H_2,GT}} \cdot T_\mathrm{tank} \cdot c_\mathrm{p,H_2}(T_\mathrm{tank}) \, + \\ 
\Big( \dot{m}_{\mathrm{H_2,FC,out}}  \cdot c_\mathrm{p,H_2}(T_\mathrm{an,HEX,out}) \, + \\ \dot{m}_{\mathrm{H_2O,FC,out}} \cdot c_\mathrm{p,H_2O}(T_\mathrm{an,HEX,out}) \Big) \cdot T_\mathrm{an,HEX,out},
	\label{eq:mcpT_mix}
	\end{split}
\end{equation}
where $c_\mathrm{p,H_2}$ and $c_\mathrm{p,H_2O}$ denote the temperature-dependent mean specific heat  capacity of hydrogen and steam, respectively, and the product $\dot{m}_\mathrm{mix} \cdot c_\mathrm{p,mix}(T_{\mathrm{mix}})$ is equal to
\begin{equation}
	\fontsize{9.9}{12}\selectfont 
	\begin{split}
\dot{m}_\mathrm{mix} \cdot c_\mathrm{p,mix}(T_{\mathrm{mix}}) = \dot{m}_{\mathrm{H_2,GT}} \cdot c_\mathrm{p,H_2}(T_{\mathrm{mix}}) \, + \\ \dot{m}_{\mathrm{H_2,FC,out}}  \cdot c_\mathrm{p,H_2}(T_{\mathrm{mix}})  +
    \dot{m}_{\mathrm{H_2O,FC,out}} \cdot c_\mathrm{p,H_2O}(T_{\mathrm{mix}}).
	\label{eq:mcp_mix}
    \end{split}
\end{equation}
\normalsize
The selected sampling plan is a Latin Hypercube Sampling scheme, which ensures that samples are evenly distributed across the input space. The outputs of interest are 
\begin{equation}
	\begin{split}
		y_\mathrm{GT} = \big[&\, P_{\mathrm{GT}}, \; \dot{m}_{\mathrm{a,FC}}, \; T_{\mathrm{HPC,out}}, \\
		&\, p_{\mathrm{HPC,out}}, \; z_{\mathrm{LPC}}, \; z_{\mathrm{HPC}}, \; T_{\mathrm{ET}}, \; N_{\mathrm{1}} \,\big]^\top, 
	\end{split}
	\label{eq:yGT}
\end{equation}
where $\dot{m}_{\mathrm{a,FC}}$ denotes the air flow from the high-pressure compressor entering the SOFC; $T_{\mathrm{HPC,out}}$ the high-pressure compressor outlet temperature; $p_{\mathrm{HPC,out}}$ the high-pressure compressor outlet pressure; $z_{\mathrm{LPC}}$ the low-pressure surge margin indicator; $z_{\mathrm{HPC}}$ the high-pressure compressor surge margin indicator; $T_{\mathrm{ET}}$ the turbine entry temperature; and  $N_{\mathrm{1}}$ the normalized spool speed. As the gas turbine exhibits nonlinear behavior that can be approximated by quadratic models, following the approach proposed by \cite{BorsboomFahdzyanaEtAl2021} and \cite{LocatelloKondaEtAl2020}, we express the outputs for all $i = 1,\dots,N_{y,\mathrm{GT}}$ as
\begin{equation}
	y_{\mathrm{GT},i} = x^\top_\mathrm{GT} Q_{\mathrm{GT},i}  x_\mathrm{GT}  + q^\top_{\mathrm{GT},i} x_\mathrm{GT} + q_{0,\mathrm{GT},i},
	\label{eq:FormulationGT}
\end{equation}
where $N_{y,\mathrm{GT}}$ represents the number of outputs, $Q_{\mathrm{GT},i}$, $q_{\mathrm{GT},i}$, and $q_{0,\mathrm{GT},i}$ are the variables including the fitting coefficients subject to identification, and $x_\mathrm{GT}$ is the vector containing models inputs, defined as
\begin{equation}
	\begin{split}
		x_\mathrm{GT} = \big[&\,  \dot{m}_{\mathrm{H_2,GT,tot}}, \; \alpha_{\mathrm{VSV,LPC}}, \; \alpha_{\mathrm{VSV,HPC}}, \; \dot{m}_{\mathrm{H_2O,FC,out}}, \; \\
		&T_{\mathrm{mix}}, \; f_{\mathrm{bleed}}, \;
		 \dot{m}_{\mathrm{a,FC,out}}, \; T_{\mathrm{cat,HEX,out}} \,\big]^\top. 
	\end{split}
	\label{eq:xGT}
\end{equation}
To identify the coefficient terms $Q_{\mathrm{GT},i}$, $q_{\mathrm{GT},i}$, and $q_{0,\mathrm{GT},i}$, we solve for all $i = 1,\dots,N_{y,\mathrm{GT}}$ the least-squares error optimization problem 
\begin{equation}
	\min_{Q_{\mathrm{GT},i},\, q_{\mathrm{GT},i},\, q_{0,\mathrm{GT},i}} \quad \Big\| y_{\mathrm{GT},i} - y_{\mathrm{GT,validated},i} \Big\|_2,
	\label{eq:least-squares-error}
\end{equation}%
\begin{table}[t] 
	\centering 
	\caption{NRMSE of the GT surrogate models} 
	\label{tab:NRMSE} 
\resizebox{\columnwidth}{!}{	\begin{tabular}{lcccccc} 
		\hline
		\centering
		\textbf{Variable} & \multicolumn{6}{c}{\textbf{NRMSE}} \\ 
		& TO 0 ft & TO 1500 ft & ToC & CRZ & FID 23000 ft & FID 1500 ft  \\  
		\hline
		$P_{\mathrm{GT}}$ & 0.87 $\%$ & 0.59 $\%$ & 0.59 $\%$ & 0.60 $\%$ & 0.90 $\%$ & 0.87 $\%$\\ 
		$\dot{m}_{\mathrm{a,FC}}$ & 0.90 $\%$ & 0.42 $\%$ & 0.44 $\%$ & 0.43 $\%$ & 0.44 $\%$ & 0.39 $\%$\\ 
		$T_{\mathrm{HPC,out}}$ & 0.29 $\%$ & 0.32 $\%$ & 0.36 $\%$ & 0.33 $\%$ & 0.30 $\%$ & 0.18 $\%$\\
		$p_{\mathrm{HPC,out}}$ & 0.93 $\%$ & 0.68 $\%$ & 0.76 $\%$ & 0.76 $\%$ & 0.80 $\%$ & 0.57 $\%$\\ 
		$z_{\mathrm{LPC}}$  & 1.67 $\%$ & 1.54 $\%$ & 1.59 $\%$ & 1.63 $\%$ & 1.54 $\%$ & 0.99 $\%$\\ 
		$z_{\mathrm{HPC}}$ & 1.46 $\%$ & 1.33 $\%$ & 1.67 $\%$ & 1.60 $\%$ & 1.57 $\%$ & 1.80 $\%$\\
		$T_{\mathrm{ET}}$ & 0.55 $\%$ & 0.44 $\%$ & 0.47 $\%$ & 0.46 $\%$ & 0.39 $\%$ & 0.19 $\%$\\
		$N_{\mathrm{1}}$ & 0.58 $\%$  & 0.50 $\%$ & 0.57 $\%$ & 0.55 $\%$ & 0.52 $\%$ & 0.29 $\%$\\
		\hline
	\end{tabular}}
\end{table}%
\noindent where $y_{\mathrm{GT,validated},i}$ denote the outputs coming from high-fidelity simulations. To improve overall modeling accuracy and capture high nonlinearities across the different operating conditions, we identify the models coefficients for the six different operating conditions described above, namely take-off at zero altitude (TO 0 ft), at 1500 ft (TO 1500 ft), top of climb (ToC), cruise (CRZ), flight idle at 23000 ft (FID 23000 ft), and at 1500ft (FID 1500 ft).   We frame the problem (\ref{eq:least-squares-error}) as a semi-definite program using the parser YALMIP \cite{Loefberg2004} and solve it using the solver MOSEK \cite{ApS2017}. The normalized root mean square error (NRMSE) values of the models for the different operating conditions are reported in Table \ref{tab:NRMSE}. The NRMSE values are normalized with respect to the corresponding maximum value of the identification dataset. Fig.~\ref{fig:GT_model_P_mf_fb} shows the comparison between the total gas turbine fuel flow reduced-order model and the high-fidelity samples. To ensure structural integrity of the rotating parts and that the high-pressure turbine blade temperature remains below a safe margin from the melting point of nickel superalloys, we limit the turbine entry temperature $T_\mathrm{ET}$ and the spool speed $N_1$ to a maximum value, thus
\begin{equation}
	T_{\mathrm{ET}} \leq T_{\mathrm{ET,max}},
	\label{eq:TET_max}
\end{equation}
\begin{equation}
	N_{\mathrm{1}} \leq N_{\mathrm{1,max}}.
	\label{eq:N1_max}
\end{equation}
Similarly, to prevent surge of the compressors, we limit the surge margin indicators to ensure a safe distance from the surge line, therefore
\begin{equation}
	z_{\mathrm{LPC}} \leq z_{\mathrm{LPC,max}},
	\label{eq:zLPC_max}
\end{equation}
\begin{equation}
	z_{\mathrm{HPC}} \leq z_{\mathrm{HPC,max}}.
	\label{eq:zHPC_max}
\end{equation}
\subsection{Heat Exchangers}
In modeling the heat exchangers, we assume that they transfer all the heat from the hot stream to the cold stream; therefore we assume to have adiabatic heat exchangers. 

\begin{figure}[t]
	\centering
	\includegraphics[width=\columnwidth]{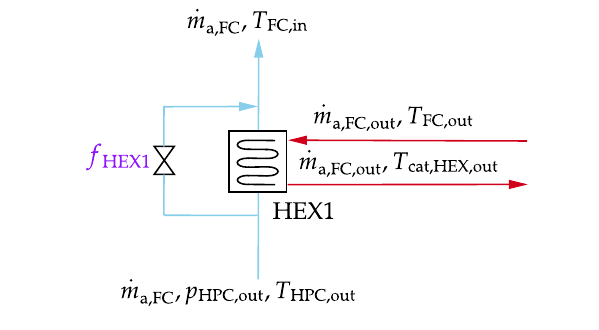}
	\caption{Simplified scheme of the air heat exchanger HEX1.}
	\label{fig:Air_HEX_scheme}
\end{figure}%
\subsubsection{\textnormal{HEX1}} 
For the air heat exchanger (Fig.~\ref{fig:Air_HEX_scheme}), which increases the temperature of the inlet SOFC air flow $\dot{m}_\mathrm{a,FC}$, the energy balance can be expressed as
\begin{equation}
	\begin{split}
		\dot{m}_\mathrm{a,HEX} \cdot  c^\mathrm{{cold}}_\mathrm{p,a}(T) \cdot \big (T^{f_\mathrm{HEX1}}_\mathrm{a,cold,out} - T_\mathrm{HPC,out} \big ) = \\ \dot{m}_\mathrm{a,FC,out} \cdot c^\mathrm{{hot}}_\mathrm{p,a}(T) \cdot \big (T_\mathrm{FC,out} - T_\mathrm{cat,HEX,out} \big ),
		\label{eq:Air_heat_exchange_f_HEX}
	\end{split}
\end{equation}
where $c^\mathrm{{cold}}_\mathrm{p,a}(T)$ denotes the temperature-dependent mean specific heat  capacity of air on the cold side, while $c^\mathrm{{hot}}_\mathrm{p,a}(T)$ refers to the temperature-dependent mean specific heat capacity of the oxygen-depleted air on the hot side. $T_\mathrm{HPC,out}$ and $T_\mathrm{FC,out}$ indicate the inlet air temperature on the cold side and the inlet oxygen-depleted air temperature on the hot side, respectively. $T^{f_\mathrm{HEX1}}_\mathrm{a,cold,out}$ and $T_\mathrm{cat,HEX,out}$ represent the outlet air temperature on the cold side and the outlet oxygen-depleted air temperature on the hot side, respectively. $	\dot{m}_\mathrm{a,HEX}$ denotes the air mass flow on the cold side that enters the HEX1, and it is equal to
\begin{equation}
	\begin{split}
		\dot{m}_\mathrm{a,HEX} = \dot{m}_\mathrm{a,FC} \cdot (1 - f_\mathrm{HEX1}).
		\label{eq:Air_HEX}
	\end{split}
\end{equation}
The resulting SOFC inlet temperature is a weighted average of the by-pass flow and the flow that enters the HEX, and can be expressed as
\begin{equation}
	T_\mathrm{a,FC,in} = 
	\frac{
		\begin{aligned}
			& f_\mathrm{HEX1}  \cdot c_\mathrm{p,a}(T) \cdot  T_\mathrm{HPC,out} \, + \\
			& (1 - f_\mathrm{HEX1})  \cdot   c_\mathrm{p,a}(T) \cdot T^{f_\mathrm{HEX1}}_\mathrm{a,cold,out}
		\end{aligned}
	}{c_\mathrm{p,a}(T)},
	\label{eq:T_mix_SOFC_in_cat}
	\end{equation}%
\noindent in which $f_\mathrm{HEX1}$ denotes the percentage of $\dot{m}_\mathrm{a,FC}$ that by-passes the air HEX. 

\emph{Design of Experiments and Surrogate Modeling Formulation}:
To construct the HEX1 surrogate model, we first perform high-fidelity evaluations within the identification space using an high-fidelity model \cite{BhapkarEtAl2024, BhapkarEtAl2025}, following a predefined sampling plan. To perform these evaluations, we first fix the design point of the heat exchanger, specifically selecting the take-off phase.  Subsequently, we perform high-fidelity evaluations and identify the models parameters through least-squares error optimization, where we aim to minimize the mean squared error between the predicted values and the high-fidelity ones. The identification space is bounded by
\begin{equation}
	\dot{m}_{\mathrm{a,HEX}} \in \left[\dot{m}^\mathrm{min}_{\mathrm{a,HEX}},  \dot{m}^\mathrm{max}_{\mathrm{a,HEX}} \right],
	\label{eq:ma_cold_HEX}
\end{equation}
\begin{equation}
	\dot{m}_{\mathrm{a,FC,out}} \in \left[\dot{m}^\mathrm{min}_{\mathrm{a,FC,out}},  \dot{m}^\mathrm{max}_{\mathrm{a,FC,out}} \right],
	\label{eq:ma_hot_HEX}
\end{equation}
\begin{equation}
	p_{\mathrm{HPC,out}} \in \left[p^\mathrm{min}_{\mathrm{HPC,out}},  p^\mathrm{max}_{\mathrm{HPC,out}} \right],
	\label{eq:p_cold_in_HEX}
\end{equation}
\begin{equation}
	T_{\mathrm{HPC,out}} \in \left[T^\mathrm{min}_{\mathrm{HPC,out}},  T^\mathrm{max}_{\mathrm{HPC,out}} \right],
	\label{eq:T_cold_in_HEX}
\end{equation}
\begin{equation}
	T_{\mathrm{FC,out}} \in \left[T^\mathrm{min}_{\mathrm{FC,out}},  T^\mathrm{max}_{\mathrm{FC,out}} \right].
	\label{eq:T_hot_in_HEX}
\end{equation}%
The selected sampling plan is a Latin Hypercube Sampling scheme. We model the effectiveness of the air heat exchanger $\varepsilon_\mathrm{a}$, defined as the ratio of the actual heat transfer to the maximum possible heat transfer, as a function of the inlet mass flow rates, temperatures, and the pressure on the cold side as
    \begin{equation}
	\varepsilon_\mathrm{a} = x^\top_\mathrm{a} Q_{\mathrm{a}}  x_\mathrm{a}  + q^\top_{\mathrm{a}} x_\mathrm{a} + q_{0,\mathrm{a}},
	\label{eq:Air_HEX_epsilon}
	\end{equation}%
	where $Q_{\mathrm{a}}$, $q_{\mathrm{a}}$, and $q_{0,\mathrm{a}}$ are the variables including the fitting coefficients identified through least-squares error optimization solving 
	\begin{equation}
\begin{aligned}
	\min_{Q_{\mathrm{a}},\, q_{\mathrm{a}},\, q_{0,\mathrm{a}}} 
	& \quad \Big\| \varepsilon_{\mathrm{a}} - \varepsilon_{\mathrm{a,validated}} \Big\|_2.
\end{aligned}
\label{eq:least-squares-error_Air_HEX}
\end{equation}%
The term $x_\mathrm{a}$ denotes the vector containing the air heat exchanger inputs, defined as
\begin{equation}
\begin{split}
	x_\mathrm{a} = \big[&\, \dot{m}_{\mathrm{a,HEX}}, \; \dot{m}_{\mathrm{a,FC,out}}, \;  p_{\mathrm{HPC,out}}, \; \\
	&\, T_{\mathrm{HPC,out}}, \; 
	T_{\mathrm{FC,out}} \, \big]^\top.
	\label{eq:x_Air_HEX_epsilon}
\end{split}
\end{equation}
The NRMSE of the effectiveness $\varepsilon_\mathrm{a}$ surrogate model, normalized by the maximum value of the training dataset, is 0.21638 $\%$. The heat transfer rate $\dot{Q}_\mathrm{a}$ transferred by the air heat exchanger, which is set equal to the enthalpy stream change, can be expressed as
\begin{equation}
\dot{Q}_\mathrm{a} = \varepsilon_\mathrm{a} \cdot \dot{Q}_\mathrm{a,max} = \Delta \dot{H}_\mathrm{a},
\label{eq:Qdot_air}
\end{equation}
where $\Delta \dot{H}_\mathrm{a}$ is given by \eqref{eq:Air_heat_exchange_f_HEX}, and $\dot{Q}_\mathrm{a,max}$ denotes the maximum possible heat trasfer rate and it is defined as
\begin{equation}
\dot{Q}_\mathrm{a,max} = C_\mathrm{a,min} \cdot (T_\mathrm{FC,out} - T_\mathrm{HPC,out}),
\label{eq:Qdot_air_max}
\end{equation}
in which the term $C_\mathrm{a,min}$ denotes the minimum heat capacity rate between the hot and cold streams.
To model the pressure loss along the path connecting the HPC outlet and the SOFC cathodic inlet, we model the outlet pressure on the cold side of the air heat exchanger as a function of the inlet conditions, therefore
\begin{equation}
p_\mathrm{a,FC,in} = x^\top_\mathrm{p,a} Q_{\mathrm{p,a}}  x_\mathrm{p,a}  + q^\top_{\mathrm{p,a}} x_\mathrm{p,a} + q_{0,\mathrm{p,a}},
\label{eq:deltap}
\end{equation}%
where the terms $Q_{\mathrm{p,a}}$, $q_{\mathrm{p,a}}$, and $q_{0,\mathrm{p,a}}$ are identified via a least-squares formulation, minimizing the error between predicted values and corresponding high-fidelity off-design model outputs \cite{MantelliEtAl2021}, and $x_{\mathrm{p,a}}$ denotes the vector 
\begin{equation}
\begin{split}
	x_\mathrm{p,a} = \big[&\, \dot{m}_{\mathrm{a,HEX}}, \; \dot{m}_{\mathrm{a,FC,out}}, \;  p_{\mathrm{HPC,out}}, \; \\
	&\, p_{\mathrm{a,FC,out}}, \; T_{\mathrm{HPC,out}}, \; 
	T_{\mathrm{FC,out}} \, \big]^\top.
	\label{eq:x_p,a}
\end{split}
\end{equation}
The NRMSE of the outlet pressure on the cold side $p_\mathrm{a,FC,in}$ surrogate model, normalized by the maximum value of the training dataset, is 0.060 $\%$.

\subsubsection{\textnormal{HEX2}} 
\begin{figure}[t]
	\centering
	\includegraphics[width=\columnwidth]{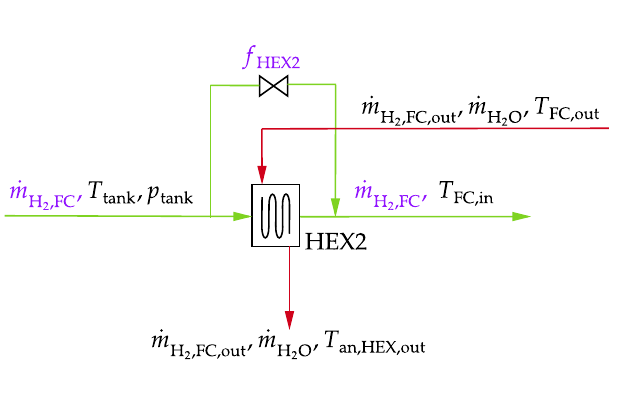}
	\caption{Simplified scheme of the fuel heat exchanger HEX2.}
	\label{fig:H2_HEX_scheme}
\end{figure}

Similarly, for the H$_2$ heat exchanger (Fig.~\ref{fig:H2_HEX_scheme}), which preheats the inlet SOFC fuel flow $\dot{m}_\mathrm{H_2,FC}$, the energy balance can be expressed as
\begin{equation}
\begin{split}
	\dot{m}_\mathrm{H_2,HEX} \cdot c^\mathrm{{cold}}_\mathrm{p,H_2}(T) \cdot \Big (T^{f_\mathrm{HEX2}}_\mathrm{H_2,cold,out} - T_\mathrm{tank} \Big ) =  \\
	\Big (\dot{m}_\mathrm{H_2,FC,out} \cdot c^\mathrm{{hot}}_\mathrm{p,H_2}(T) \, +  \\
	\dot{m}_\mathrm{H_2O} \cdot c^\mathrm{{hot}}_\mathrm{p,H_2O}(T) \Big ) \cdot \Big (T_\mathrm{FC,out} \, - 
	T_\mathrm{an,HEX,out} \Big ),
\end{split}
\label{eq:H2_heat_exchange_fHEX}
\end{equation}
where $c^\mathrm{{cold}}_\mathrm{p,H_2}(T)$ and $c^\mathrm{{hot}}_\mathrm{p,H_2}(T)$ denote the mean specific heat capacity of hydrogen on the cold and hot sides, respectively, and $c^\mathrm{{cold}}_\mathrm{p,H_2O}(T)$ and $c^\mathrm{{hot}}_\mathrm{p,H_2O}(T)$ denote the mean specific heat capacity of steam on the cold and hot sides, respectively, and where $\dot{m}_\mathrm{H_2,HEX}$ denotes the portion of SOFC inlet hydrogen flow $\dot{m}_\mathrm{H_2,FC}$ that enters the heat exchanger HEX2, and it is equal to
\begin{equation}
\begin{split}
	\dot{m}_\mathrm{H_2,HEX} = \dot{m}_\mathrm{H_2,FC} \cdot (1 - f_\mathrm{HEX2}).
	\label{eq:H2_flow_HEX}
\end{split}
\end{equation}
$T_\mathrm{tank}$ and $T_\mathrm{FC,out}$ denote the inlet H$_2$ temperature on the cold side and the inlet anode off-gases temperature on the hot side, respectively. $T^{f_\mathrm{HEX2}}_\mathrm{H_2,cold,out}$ and $T_\mathrm{an,HEX,out}$ denote the outlet H$_2$ temperature on the cold side and the outlet anode off-gases temperature on the hot side, respectively. The resulting SOFC inlet temperature is a weighted average of the by-pass flow and the flow that enters the HEX, and can be expressed as
\begin{equation}
		T_\mathrm{H_2,FC,in} =
		\frac{
			\begin{aligned}
				& f_\mathrm{HEX2} \cdot c_\mathrm{p,H_2}(T) \cdot T_\mathrm{tank} \, + \\
				& (1 - f_\mathrm{HEX2}) \cdot c_\mathrm{p,H_2}(T) \cdot T^{f_\mathrm{HEX2}}_\mathrm{H_2,cold,out}
			\end{aligned}
		}{
			c_\mathrm{p,H_2}(T)
		}
		\label{eq:T_mix_SOFC_in_an}
\end{equation}%
\noindent in which the fraction $f_\mathrm{HEX2}$ denotes percentage of $\dot{m}_\mathrm{H_2,FC}$ that by-passes the H$_2$ HEX. 

\emph{Design of Experiments and Surrogate Modeling Formulation}: To construct the HEX2 surrogate model, we first perform high-fidelity evaluations within the identification space using an high-fidelity model, following a predefined sampling scheme. To perform these evaluations, we first fix the design point of the heat exchanger, specifically for the take-off phase. Subsequently, we perform high-fidelity evaluations and identify the models parameters via least-squares minimization. The identification space is bounded by
\begin{equation}
	\dot{m}_{\mathrm{H_2,HEX}} \in \left[\dot{m}^\mathrm{min}_{\mathrm{H_2,HEX}},  \dot{m}^\mathrm{max}_{\mathrm{H_2,HEX}} \right],
	\label{eq:mH2_cold_HEX}
\end{equation}
\begin{equation}
	\dot{m}_{\mathrm{H_2,FC,out}} \in \left[\dot{m}^\mathrm{min}_{\mathrm{H_2,FC,out}},  \dot{m}^\mathrm{max}_{\mathrm{H_2,FC,out}} \right],
	\label{eq:mH2_hot_HEX}
\end{equation}
\begin{equation}
	\dot{m}_{\mathrm{H_2O,FC,out}} \in \left[\dot{m}^\mathrm{min}_{\mathrm{H_2O,FC,out}},  \dot{m}^\mathrm{max}_{\mathrm{H_2O,FC,out}} \right],
	\label{eq:mH2O_hot_HEX}
\end{equation}
\begin{equation}
	p_{\mathrm{tank}} \in \left[p^\mathrm{min}_{\mathrm{tank}},  p^\mathrm{max}_{\mathrm{tank}} \right],
	\label{eq:p_cold_in_H2_HEX}
\end{equation}
\begin{equation}
	T_{\mathrm{FC,out}} \in \left[T^\mathrm{min}_{\mathrm{FC,out}},  T^\mathrm{max}_{\mathrm{FC,out}} \right].
	\label{eq:T_hot_in_H2_HEX}
\end{equation}
The selected sampling plan is a Latin Hypercube Sampling scheme. We model the effectiveness of the H$_2$ HEX as a function of the inlet mass flows, temperatures, and the pressure on the cold side as
\begin{equation}
\varepsilon_\mathrm{H_2} = x^\top_\mathrm{H_2} Q_{\mathrm{H_2}}  x_\mathrm{H_2}  + q^\top_{\mathrm{H_2}} x_\mathrm{H_2} + q_{0,\mathrm{H_2}},
\label{eq:H2_HEX_epsilon}
\end{equation}%
where $Q_{\mathrm{H_2}}$, $q_{\mathrm{H_2}}$, and $q_{0,\mathrm{H_2}}$ are identified through least-squares error optimization solving 
\begin{equation}
\begin{aligned}
\min_{Q_{\mathrm{H_2}},\, q_{\mathrm{H_2}},\, q_{0,\mathrm{H_2}}} & \quad \Big\| \varepsilon_{\mathrm{H_2}} - \varepsilon_{\mathrm{H_2,validated}} \Big\|_2.
\label{eq:least-squares-error_H2_HEX}
\end{aligned}
\end{equation}%
The term $x_\mathrm{H_2}$ denotes the vector containing the H$_2$ heat exchanger inputs, defined as
\begin{equation}
\begin{split}
x_\mathrm{H_2} = \big[&\, \dot{m}_{\mathrm{H_2,HEX}}, \; \dot{m}_{\mathrm{H_2,FC,out}},
\dot{m}_{\mathrm{H_2O,FC,out}}, \\
&\, \;  p_{\mathrm{tank}}, \;   
T_{\mathrm{FC,out}} \, \big]^\top.
\label{eq:x_H2_HEX_epsilon}
\end{split}
\end{equation}
The NRMSE of the effectiveness $\varepsilon_\mathrm{H_2}$ surrogate model, normalized by the maximum value of the training dataset, is 0.967 $\%$. The heat transfer rate $\dot{Q}_\mathrm{H_2}$ transferred by the H$_2$ heat exchanger, which is set equal to the enthalpy stream change, can be expressed as
\begin{equation}
\dot{Q}_\mathrm{H_2} = \varepsilon_\mathrm{H_2} \cdot \dot{Q}_\mathrm{H_2,max} = \Delta \dot{H}_\mathrm{H_2},
\label{eq:Qdot_H2}
\end{equation}
where $\Delta \dot{H}_\mathrm{H_2}$ is given by \eqref{eq:H2_heat_exchange_fHEX}, and $\dot{Q}_\mathrm{H_2,max}$ denotes the maximum possible heat trasfer rate and it is defined as
\begin{equation}
\dot{Q}_\mathrm{H_2,max} = C_\mathrm{H_2,min} \cdot (T_\mathrm{FC,out} - T_\mathrm{tank}).
\label{eq:Qdot_H2_max}
\end{equation}
The term $C_\mathrm{H_2,min}$ denotes the minimum heat capacity rate between the hot and cold streams. Similarly to (\ref{eq:deltap}), we model the outlet pressure on the cold side of the H$_2$ heat exchanger as a function of the inlet conditions, therefore
\begin{equation}
	p_\mathrm{H_2,FC,in} = x^\top_\mathrm{p,H_2} Q_{\mathrm{p,H_2}}  x_\mathrm{p,H_2}  + q^\top_{\mathrm{p,H_2}} x_\mathrm{p,H_2} + q_{0,\mathrm{p,H_2}},
	\label{eq:deltap_H2}
\end{equation}
where the terms $Q_{\mathrm{p,H_2}}$, $q_{\mathrm{p,H_2}}$, and $q_{0,\mathrm{p,H_2}}$ are identified via a least-squares formulation, minimizing the error between predicted values and corresponding high-fidelity off-design model outputs, and where denotes the vector
\begin{equation}
	\begin{split}
x_\mathrm{H_2} = \big[&\, \dot{m}_{\mathrm{H_2,HEX}}, \; \dot{m}_{\mathrm{H_2,FC,out}},
\dot{m}_{\mathrm{H_2O,FC,out}}, \\
&\, \;  p_{\mathrm{tank}}, \; p_{\mathrm{H_2,FC,out}}, \;  
T_{\mathrm{tank}}, \; T_{\mathrm{FC,out}} \, \big]^\top.
		\label{eq:x_p,H2}
	\end{split}
\end{equation}
The NRMSE of the outlet pressure on the cold side $p_\mathrm{H_2,FC,in}$ surrogate model, normalized by the maximum value of the training dataset, is 0.086 $\%$. 
We assume, for simplicity, to have the same pressure and temperature on the anode and cathode side, therefore
\begin{equation}
	p_\mathrm{H_2,FC,in} =p_\mathrm{a,FC,in} = 	p_\mathrm{FC,in}.
	\label{eq:deltap_equivalence}
\end{equation}
and
\begin{equation}
	T_\mathrm{H_2,FC,in} = T_\mathrm{a,FC,in} = T_\mathrm{FC,in}.
	\label{eq:deltaT_equivalence}
\end{equation}
Furthermore, as the cell geometry is not known, the pressure drop across the stack cannot be estimated reliably and, therefore, it is neglected. Consequently, 
\begin{equation}
	p_\mathrm{H_2,FC,out} = p_\mathrm{a,FC,out} = 	p_\mathrm{FC,in}.
	\label{eq:dp_zero}
\end{equation} 
\subsection{Solid Oxide Fuel Cell}
	\begin{figure}[]
	\centering
	\includegraphics[width=\columnwidth]{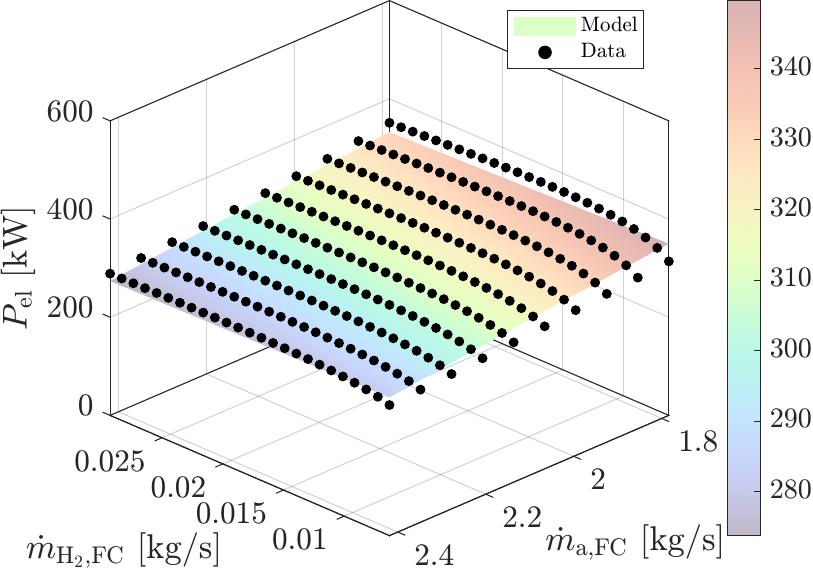}
	\caption{Comparison between the SOFC electrical power reduced-order model and the high-fidelity samples as function of the inlet hydrogen flow $\dot{m}_{\mathrm{H_2,FC}}$ and the inlet air flow $\dot{m}_\mathrm{a,FC}$ for constant inlet temperature $T_\mathrm{FC,in}$, inlet pressure $p_\mathrm{FC,in}$, and voltage $U_\mathrm{FC}$.}
	\label{fig:PWcell}
\end{figure}
\begin{figure}[t]
	\centering
	\includegraphics[width=\columnwidth]{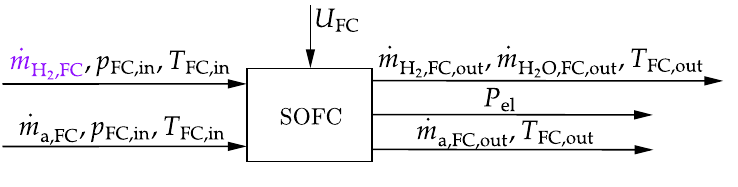}
	\caption{Simplified scheme representing the SOFC interfaces. The SOFC receives from the GT an air flow $\dot{m}_\mathrm{a,FC}$ at temperature $T_\mathrm{FC,in}$ and pressure $p_\mathrm{FC,in}$. An hydrogen flow $\dot{m}_\mathrm{H_2,FC}$ is injected from the storage tank and reaches the SOFC inlet at temperature $T_\mathrm{FC,in}$ and pressure $p_\mathrm{FC,in}$. Additionally, a voltage $U_\mathrm{FC}$ is applied to the SOFC. As a result, the SOFC produces an electrical power $P_\mathrm{el}$. The exhaust gases, $\dot{m}_\mathrm{H_2,FC,out}$, $\dot{m}_\mathrm{H_2O,FC,out}$, and $\dot{m}_\mathrm{a,FC,out}$, leave the SOFC at temperature $T_\mathrm{FC,out}$.}
	\label{fig:SOFC_interfaces}
\end{figure}

%
In this section, we present the models of the SOFC component and their identification. Based on the block diagram in Fig.~\ref{fig:SOFC_interfaces}, we first evaluate high-fidelity samples, from which we construct the surrogate reduced-order models.

\emph{Design of Experiments and Surrogate Modeling Formulation}: To construct the SOFC surrogate models, we perform high-fidelity evaluations within the identification space using a 1D SOFC performance model, developed based on the work of \cite{LeonideEtAl2009} and \cite{KlotzEtAl2014}, which has been validated under atmospheric operating conditions. The parameters of the reduced-order models are then identified through least-squares error optimization, where we aim to minimize the mean squared error between the predicted values and the high-fidelity ones. 
The identification space for the high-fidelity evaluations is bounded by 
\begin{equation}
	p_{\mathrm{FC,in}} \in \left[p^\mathrm{min}_{\mathrm{FC,in}},  p^\mathrm{max}_{\mathrm{FC,in}} \right],
	\label{eq:pFC}
\end{equation}
\begin{equation}
	T_{\mathrm{FC,in}} \in \left[T^\mathrm{min}_{\mathrm{FC,in}},  T^\mathrm{max}_{\mathrm{FC,in}} \right],
	\label{eq:TFC}
\end{equation}
\begin{equation}
	U_\mathrm{FC} \in \left[U^\mathrm{min}_{\mathrm{FC}},  U^\mathrm{max}_{\mathrm{FC}} \right],
	\label{eq:U}
\end{equation}
\begin{equation}
	\dot{m}_\mathrm{H_2,FC} \in \left[\dot{m}^\mathrm{min}_{\mathrm{H_2,FC}},  \dot{m}^\mathrm{max}_{\mathrm{H_2,FC}} \right],
	\label{eq:mfFC}
\end{equation}
\begin{equation}
	\dot{m}_\mathrm{a,FC} \in \left[\dot{m}^\mathrm{min}_{\mathrm{a,FC}},  \dot{m}^\mathrm{max}_{\mathrm{a,FC}} \right].
	\label{eq:maFC}
\end{equation}
The selected sampling plan is a Latin Hypercube Sampling scheme. The outputs of interest are 
\begin{equation}
	\begin{split}
		y_\mathrm{FC} = \big[&\, P_{\mathrm{el}}, \; T_{\mathrm{FC,out}}, \; \dot{m}_{\mathrm{H_2,FC,out}}, \;  \dot{m}_{\mathrm{a,FC,out}} \,\big]^\top, 
	\end{split}
	\label{eq:yFC}
\end{equation}
where $P_\mathrm{el}$ denotes the total electrical power produced by the SOFC, and it is equal to
\begin{equation}
P_\mathrm{el} = P_\mathrm{density,cell} \cdot A_\mathrm{act} \cdot N_\mathrm{cells},
\label{eq:Pel_SOFC}
\end{equation}
where $P_\mathrm{density,cell}$ denotes the single cell area-specific power density, $A_\mathrm{act}$ the active area, and $N_\mathrm{cells}$ the total number of cells. $T_{\mathrm{FC,out}}$ denotes the temperature of the SOFC off-gases, $\dot{m}_{\mathrm{H_2,FC,out}}$ denotes the unutilized hydrogen flow coming out from the SOFC, and $\dot{m}_{\mathrm{a,FC,out}}$ denotes the oxygen-depleted air. Similarly to \eqref{eq:FormulationGT}, we express for all $i = 1,\dots,N_{y,\mathrm{FC}}$ the outputs of interest as
\begin{equation}
	y_{\mathrm{FC},i} = x^\top_\mathrm{FC} Q_{\mathrm{FC},i}  x_\mathrm{FC}  + q^\top_{\mathrm{FC},i} x_\mathrm{FC} + q_{0,\mathrm{FC},i},
	\label{eq:FormulationFC}
\end{equation}
where $N_{y,\mathrm{FC}}$ is the total number of outputs, $Q_{\mathrm{FC},i}$, $q_{\mathrm{FC},i}$, and $q_{0,\mathrm{FC},i}$ are the variables including the fitting coefficients to be identified, and $x_\mathrm{FC}$ is the vector containing models inputs, defined as
\begin{equation}
	\begin{split}
		x_\mathrm{FC} = \big[&\, p_{\mathrm{FC,in}}, \; T_{\mathrm{FC,in}}, \; U_{\mathrm{FC}}, \; \dot{m}_{\mathrm{H_2,FC}}, \; \dot{m}_{\mathrm{a,FC}} \,\big]^\top. 
	\end{split}
	\label{eq:xFC}
\end{equation}
To identify the coefficient terms $Q_{\mathrm{FC},i}$, $q_{\mathrm{FC},i}$, and $q_{0,\mathrm{FC},i}$, we frame for all $i = 1,\dots,N_{y,\mathrm{FC}}$ the least-squares error optimization problem 
\begin{equation}
	\min_{Q_{\mathrm{FC},i},\, q_{\mathrm{FC},i},\, q_{0,\mathrm{FC},i}} \quad \Big\| y_{\mathrm{FC},i} - y_{\mathrm{FC,validated},i} \Big\|_2,
	\label{eq:least-squares-error_FC}
\end{equation}
in which where $y_{\mathrm{FC,validated},i}$ denote the outputs coming from high-fidelity simulations, as a semi-definite program using the parser YALMIP and solve it using the solver MOSEK. 
\begin{table}[] 
	\centering 
	\caption{NRMSE of the SOFC surrogate models} 
	\label{tab:NRMSE_SOFC} 
	\begin{tabular}{lc} 
		\hline
		\centering
		\textbf{Variable} & \textbf{NRMSE} \\
		\hline
		$P_{\mathrm{el}}$  & 1.94 $\%$ \\ 
		$T_{\mathrm{FC,out}}$  & 0.67 $\%$ \\ 
		$\dot{m}_{\mathrm{H_2,FC,out}}$ & 0.042 $\%$ \\
		$\dot{m}_{\mathrm{a,FC,out}}$ & 0.036 $\%$\\
		\hline
	\end{tabular} 
\end{table} 
The NRMSE values with respect to the maximum value of the SOFC surrogate models are reported in Table~\ref{tab:NRMSE_SOFC}.  Fig.~\ref{fig:PWcell} shows the comparison between the electrical power reduced-order model and the high-fidelity samples for constant inlet temperature $T_\mathrm{FC,in}$, inlet pressure $p_\mathrm{FC,in}$, and voltage $U_\mathrm{FC}$, confirming the good agreement between the model predictions and the validated data. 
Based on the validated reduced-order model, we can calculate key operating quantities of the SOFC, such as the outlet steam $\dot{m}_\mathrm{H_2O,FC,out}$ and both the air and fuel utilization factors. 
In particular, we obtain the outlet steam $\dot{m}_\mathrm{H_2O,FC,out}$ as
\begin{equation}
	\dot{m}_\mathrm{H_2O,FC,out} = \Big ( \dot{m}_\mathrm{H_2,FC} - \dot{m}_\mathrm{H_2,FC,out} \Big ) \cdot \frac{M_\mathrm{H_2O}}{M_\mathrm{H_2}},
	\label{eq:Steam}
\end{equation}
where $M_\mathrm{H_2O}$ and $M_\mathrm{H_2}$ denote the molar weight of the steam and hydrogen, respectively, $\dot{m}_\mathrm{H_2O,FC}$ denotes the inlet steam flow entering the SOFC, while the second term on the right-hand side accounts for the steam produced inside the cell by the electrochemical reaction. In addition, the fuel utilization factor $f_\mathrm{FU}$ is given by
\begin{equation}
	f_\mathrm{FU} = \frac{\dot{m}_\mathrm{H_2,FC} - \dot{m}_\mathrm{H_2,FC,out}}{\dot{m}_\mathrm{H_2,FC}},
	\label{eq:Uf}
\end{equation}
and represents the ratio of the total hydrogen consumed and the hydrogen supplied, while the air utilization factor is given by
\begin{equation}
	f_\mathrm{AU} = \frac{\dot{m}_\mathrm{a,FC} - \dot{m}_\mathrm{a,FC,out}}{\dot{m}_\mathrm{O_2,FC}},
	\label{eq:Ua}
\end{equation}
and represents the ratio of the total oxygen consumed and the oxygen supplied. To prevent damage and maintain safe operation, we limit the fuel utilization factor and the temperature difference across the SOFC, denoted by $\Delta T,$ to a maximum value, therefore
\begin{equation}
	f_\mathrm{FU} \leq f_\mathrm{FU,max},
	\label{eq:Uf_max}
\end{equation}
\begin{equation}
	T_\mathrm{FC,out} - T_\mathrm{FC,in}  \leq \Delta T_\mathrm{max}.
	\label{eq:DeltaT_max}
\end{equation}

\subsection{Battery}
In this section, we derive a model for the battery pack. We model the battery output power $P_\mathrm{b}$ as 
\begin{equation}
	P_\mathrm{b} = P_\mathrm{i} - R \cdot \frac{P^2 _\mathrm{i}}{U^2_\mathrm{oc}},
	\label{eq:Pi}
\end{equation}
where $P_\mathrm{i}$ denotes the internal battery power, $R$ denotes the battery pack resistance, and $U_\mathrm{oc}$ denotes the open-circuit voltage. The battery state of energy $E_\mathrm{b}$ changes as function of $P_\mathrm{i}$ as
\begin{equation}
	\frac{\mathrm{d}}{\mathrm{d}t} E_\mathrm{b} = -P_\mathrm{i}.
	\label{eq:Eb}
\end{equation}
We bound the battery state of energy with the minimum and the maximum state-of-charge limits $\xi^{\mathrm{min}}_\mathrm{b}$ and $\xi^{\mathrm{max}}_\mathrm{b}$ as 
\begin{equation}
	E_\mathrm{b} \in \left [\xi^{\mathrm{min}}_\mathrm{b}, \xi^{\mathrm{max}}_\mathrm{b} \right ] \cdot  E_\mathrm{b,max},
	\label{eq:Eb_limits}
\end{equation}
where $E_\mathrm{b,max}$ is the total battery energy capacity, and $\xi^{\mathrm{min}}_\mathrm{b}$, and $\xi^{\mathrm{max}}_\mathrm{b}$ denote the minimum and maximum state-of-charge levels, respectively. We assume that the battery is fully charged at the beginning of the flight cycle, and therefore
\begin{equation}
	E_\mathrm{b}(0) = \xi^{\mathrm{max}}_\mathrm{b} \cdot  E_\mathrm{b,max}.
	\label{eq:Eb0}
\end{equation}
To model the additional weight of the batteries, one per propulsion system, with a certain battery specific energy density $e_\mathrm{b}$, we assume to have a steady, constant speed cruise flight, modeling the aircraft as a point mass. Therefore, the lift $L$ equals the weight $W$, and the required propulsive power can be expressed as product between the drag force and the speed as
\begin{equation}
	P_\mathrm{req,prop} = \left ( \frac{1}{2} \cdot \rho \cdot S \cdot C_\mathrm{D} \cdot v^2 \right ) \cdot v,
	\label{eq:Preq_D}
\end{equation}
where $\rho$ denotes the air density, $S$ the wing surface area, $C_\mathrm{D}$ the drag coefficient, and $v$ the speed. The drag coefficient $C_\mathrm{D}$ can be expressed using the parabolic drag polar model
\begin{equation}
	 C_\mathrm{D} =  C_\mathrm{D0} +  k \cdot C^2_\mathrm{L},
	\label{eq:Cd}
\end{equation}
in which $C_\mathrm{D0}$ denotes the parasitic drag coefficient, and $k$ the induced drag factor, and by substituting the lift coefficient $C_\mathrm{L}$ 
\begin{equation}
	C_\mathrm{L} =  \frac{2\cdot W}{\rho \cdot S \cdot v^2}
	\label{eq:CL}
\end{equation}
in \eqref{eq:Cd}, the required propulsive power becomes equal to
\begin{equation}
	P_\mathrm{req,prop} = \frac{1}{2} \cdot \rho \cdot S \cdot C_\mathrm{D0} \cdot v^3 + \frac{2 \cdot k}{\rho \cdot S \cdot v} \cdot W^2.
	\label{eq:Preq_W}
\end{equation}
Since only the induced-drag term depends on the aircraft weight, the increase in required propulsive power due to the battery mass can be approximated by expanding the weight-dependent term in \eqref{eq:Preq_W} around the nominal aircraft weight $W_0$. Under these conditions, the increase of the required propulsive power due to the additional batteries weight can be expressed as
\begin{equation}
	\fontsize{9.48}{12}\selectfont 
	P_\mathrm{req,prop,b} = P_\mathrm{req,prop} \cdot  \Big ( 1 + \lambda \cdot \Big ( \Delta W_\mathrm{b}^2 + 2 \cdot W_0 \cdot \Delta W_\mathrm{b}  \Big ) \Big ),
	\label{eq:Preq_bat}
\end{equation}
where $P_\mathrm{req,prop,b}$ denotes the required propulsive power corresponding to the case with the battery, $P_\mathrm{req,prop}$ is the reference requested power without the battery, $W_0$ is the initial weight of the aircraft, and $\Delta W_\mathrm{b}$ denotes the additional weight from the two batteries. The coefficient $\lambda$ accounts for aerodynamic characteristics, flight altitude, wing area, cruise speed, and the initial aircraft weight.
For the considered aircraft,  $\lambda$ is assumed equal to 1.08 $\times$ 10$^{-10}$ N$^{-2}$. Although derived under steady cruise conditions, we assume that the same power scaling law applies to the remaining mission phases.
\subsection{Minimum-fuel Optimization Problem}
Summarizing the objective function, the component models, and the constraints derived in the previous sections, we can formulate the minimum-fuel optimization problems for hydrogen-based hybrid SOFC-GT aero engines with and without the battery pack, for the cases with and without heat exchangers by-pass valves. The optimization is framed as a multi-point steady-state problem, in which each operating point in Fig.~\ref{fig:Mission} is treated as an independent steady-state point. The total fuel consumption over the mission is approximated by a weighted sum of the fuel consumption at each operating point, where the weights correspond to the duration of each flight phase. We state the minimum-fuel optimization problem for the case without the battery as:
\begin{prob}[GT-SOFC Minimum-fuel Optimization Problem]
	\label{prob:Prob1} 
	\begin{equation*}
		\begin{aligned} 
			&\!\min_{\substack{
					\mathbf{u}\\
			}} & & \int_{0}^{t_\mathrm{fin}}{\dot{m}_{\mathrm{H_2,tot}}(t)} \mathrm{d} t, \\ & \textnormal{s.t. } & &P_\mathrm{req,prop}(t) = P_\mathrm{GT}(t) + P_\mathrm{EM}(t),\\ 
			& & &P_\mathrm{EM}(t) = \left (P_\mathrm{el}(t) - P_\mathrm{aux}(t) \right ) \cdot \eta_\mathrm{el}, \\ 
			& & &\eqref{eq: total fuel}-\eqref{eq:xGT}, \eqref{eq:TET_max} - \eqref{eq:Air_HEX_epsilon}, \\ & & & \eqref{eq:x_Air_HEX_epsilon} - \eqref{eq:H2_HEX_epsilon},  \eqref{eq:x_H2_HEX_epsilon} - \eqref{eq:xFC}, \\
			& & & \eqref{eq:Steam} - \eqref{eq:DeltaT_max}, \\
			\end{aligned} 
	\end{equation*} 
\end{prob}
\noindent where $\mathbf{u}$ denotes the inputs vector
\begin{equation}
	\begin{split}
   \mathbf{u} = \big [&\, \dot{m}_{\mathrm{H_2,GT}}(t), \;
   \dot{m}_{\mathrm{H_2,FC}}(t), \; U_{\mathrm{FC}}(t), \;  \alpha_{\mathrm{VSV,LPC}}(t), \; \\
   &\, \alpha_{\mathrm{VSV,HPC}}(t), \;
   f_\mathrm{bleed}(t), \; f_\mathrm{HEX1}(t), \; f_\mathrm{HEX2}(t) \; \big ],
    \end{split}
	\label{eq:u_inputs}
\end{equation}
$P_\mathrm{EM}$ denotes the mechanical power produced by the electric motor, $P_\mathrm{aux}$ the auxiliary electrical power demand, and $\eta_\mathrm{el}$ the overall electrical efficiency, accounting for the efficiencies of cables, converter, inverter, and the electric machine. In contrast, for the configuration including the battery, a discrete-time dynamic coupling is introduced through the battery state of energy. Therefore, we frame the minimum-fuel optimization problem for the hydrogen-based GT-SOFC hybrid propulsion system integrating a battery as a multi-point optimization problem with discrete-time dynamics as:\\
\needspace{6\baselineskip}
\begin{prob}[GT-SOFC-Battery Minimum-fuel Optimization Problem]
	\label{prob:Prob2} 
	\begin{equation*} 
		\begin{aligned} 
			&\!\min_{\substack{
					\bm{\mathrm{u_b}}\\
			}} & & \int_{0}^{t_\mathrm{fin}}{\dot{m}_{\mathrm{H_2,tot}}(t)} \mathrm{d} t, \\ & \textnormal{s.t. } & &P_\mathrm{req,prop,b}(t) = P_\mathrm{GT}(t) + P_\mathrm{EM}(t),\\ 
			& & &P_\mathrm{EM}(t) = \left (P_\mathrm{el}(t) - P_\mathrm{aux}(t) + P_\mathrm{b}(t) \right ) \cdot \eta_\mathrm{el}, \\ 
				& & &\eqref{eq: total fuel}-\eqref{eq:xGT}, \eqref{eq:TET_max} - \eqref{eq:Air_HEX_epsilon}, \\ & & & \eqref{eq:x_Air_HEX_epsilon} - \eqref{eq:H2_HEX_epsilon},  \eqref{eq:x_H2_HEX_epsilon} - \eqref{eq:xFC}, \\
			& & & \eqref{eq:Steam} - \eqref{eq:Preq_bat}, \\ 
			\end{aligned} 
	\end{equation*} 
\end{prob}
 where $\bm{\mathrm{u_b}}$ denotes the inputs vector
\begin{equation}
	\begin{split}
		\bm{\mathrm{u_b}} = \big [&\, \dot{m}_{\mathrm{H_2,GT}}(t), \;
		\dot{m}_{\mathrm{H_2,FC}}(t), \; U_{\mathrm{FC}}(t), \;  \alpha_{\mathrm{VSV,LPC}}(t), \;  P_\mathrm{b}(t) \;  \\
		&\, \alpha_{\mathrm{VSV,HPC}}(t), \;
		f_\mathrm{bleed}(t), \; f_\mathrm{HEX1}(t),\; f_\mathrm{HEX2}(t)\;\big ].
	\end{split}
	\label{eq:ub_inputs}
\end{equation}
\subsection{Discussion}
A few comments are in order. First, we neglect the additional weight of pipes and valves associated with the by-pass valves configurations. Second, we assume that the pressure drop between the high-pressure compressor (HPC) outlet and the combustion chamber inlet is set to be the same as that along the path connecting the HPC outlet, passing through the SOFC, and returning to the combustion chamber.
Third, we assume that the fuel system can adjust the pressure of the fuel entering the HEX2 $p_\mathrm{tank}$ to match the pressure on the cathode side, as stated in (\ref{eq:deltap_equivalence}).
Fourth, we assume that all cells in the stack operate under the same conditions and deliver uniform performance in line with~\cite{SapraEtAl2021}.
Finally, we assume to have adiabatic conditions along the paths connecting the SOFC and the gas turbine; therefore, any heat exchange with the ambient environment is neglected. 

%% file: Sections/results.tex
\section{Optimization Results}
In this section, we validate our approach against high-fidelity simulations results and present the  trajectories obtained by solving both Problem~\ref{prob:Prob1} and Problem~\ref{prob:Prob2}.
Furthermore, we analyze and compare the optimal operating strategies and total fuel consumption of the reference system with those of the same configurations integrating heat exchangers by-pass valves. Table~\ref{tab:GT_Parameters} and Table~\ref{tab:SOFC_Parameters} show the GT parameters and the thermodynamic and SOFC parameters, respectively, used in our analysis. Table~\ref{tab:Mission_Parameters} shows mission dependent parameters, in which $\dot{m}_\mathrm{aux}$ denotes the auxiliary engine bleed, while $\Delta T_\mathrm{ISA}$ the temperature deviation compared to the International Standard Atmosphere (ISA). Table~\ref{tab:Battery_Parameters} shows the battery parameters. We optimize the operation of the four propulsion configurations over a given power profile. We parse the two problems with CasADi \cite{AnderssonGillisEtAl2019}, for the cases with and without by-pass valves, and solve them with the nonlinear solver IPOPT \cite{WachterBiegler2006}. For each of the optimization problems, the parsing takes less than 2 seconds, while the solving takes less than 3 seconds, on a system with a 12th Gen Intel Core i7 processor and 16 GB RAM.
\begin{table}[htbp] 
	\centering 
	\caption{GT parameters} 
	\label{tab:GT_Parameters} 
	\begin{tabular}{lcc} 
		\hline
		\centering
		\textbf{Parameter} & \textbf{Value} & \textbf{Unit} \\ 
		\hline
		$T_\mathrm{ET,max}$ & 1500 & [K] \\
		$N_\mathrm{1,max}$ & 1 & [-] \\
		$z_\mathrm{LPC,max}$ & 0.98 & [-] \\
		$z_\mathrm{LPC,max}$ & 0.98 & [-] \\
		\hline
	\end{tabular} 
\end{table}

\begin{table}[htbp] 
	\centering 
	\caption{SOFC, electrical, and thermodynamic parameters} 
	\label{tab:SOFC_Parameters} 
	\begin{tabular}{lcc} 
		\hline
		\centering
		\textbf{Parameter} & \textbf{Value} & \textbf{Unit} \\ 
		\hline
		$N_\mathrm{cells}$ & 1048 & [-] \\
		$A_\mathrm{act}$ & 103.43 & [cm$^2$] \\
		$f_\mathrm{FU,max}$ & 0.6 & [-] \\
		$\Delta T_\mathrm{max}$ & 200 & [K] \\
		$\eta_\mathrm{el}$ & 0.92 & [-] \\
		$T_\mathrm{tank}$ & 300 & [K] \\
		\hline
	\end{tabular} 
\end{table}
\begin{table}[htbp]
	\centering
	\caption{Mission-dependent parameters}
	\label{tab:Mission_Parameters}
	\resizebox{\columnwidth}{!}{\begin{tabular}{lccccccc}
		\hline
		\textbf{Condition} & $M$ [-] & $\dot{m}_\mathrm{aux}$  [kg/s] & $\Delta T_\mathrm{ISA}$  [K] & $P_\mathrm{req,prop}$ [kW]& $P_\mathrm{aux}$ [kW] \\
		\hline
		TO 0 ft     & 0.168 & 0    & 0  & 1390 & 25   \\
		TO 1500 ft  & 0.218 & 0.13 & 0  & 1390 & 25 \\
		ToC         & 0.33  & 0.13 & 10 & 580 & 22.5 \\
		CRZ         & 0.4   & 0.13 & 0  & 560 & 22.5 \\
		FID 23000 ft& 0.468 & 0.13 & 0  & 295 & 22.5 \\
		FID 1500 ft & 0.310 & 0.13 & 0  & 295 & 22.5 \\
		\hline
	\end{tabular}}
\end{table}%

\begin{table}[htbp] 
	\centering 
	\caption{Battery parameters} 
	\label{tab:Battery_Parameters} 
	\begin{tabular}{lcc} 
		\hline
		\centering
		\textbf{Parameter} & \textbf{Value} 
		& \textbf{Unit} \\
		\hline
		$U_\mathrm{oc}$ & 324 & [V]  \\
		$R$ & 33.75 & [m$\mathrm{\Omega}$]  \\
		$\xi^{\mathrm{min}}_\mathrm{b}$ & 0.2 & [-]  \\
		$\xi^{\mathrm{max}}_\mathrm{b}$ & 0.9 & [-]  \\
		$E_\mathrm{b,max}$ & 25.92 & [kWh] \\
		$e_\mathrm{b}$  & 0.4 & [kWh/kg]  \\
		\hline		
	\end{tabular} 
\end{table}
\begin{figure}[htbp]
	\centering
	\includegraphics[width=\linewidth]{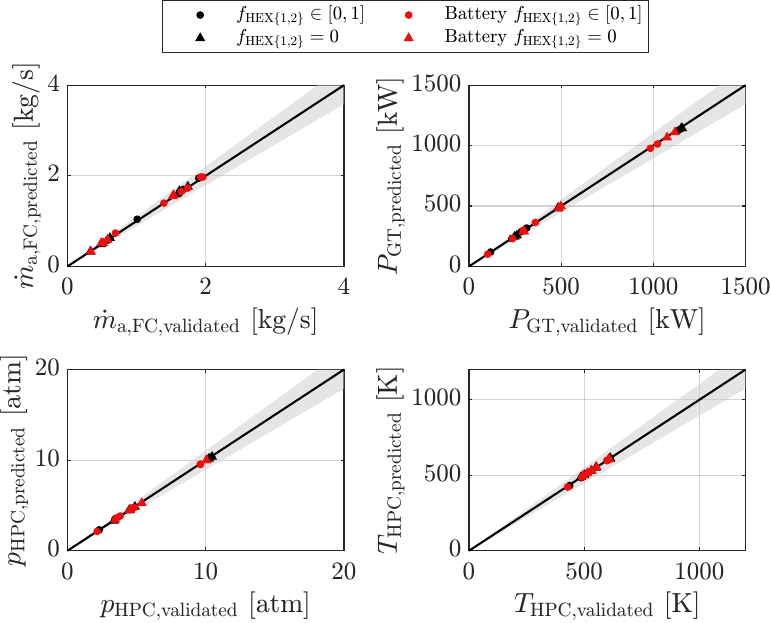}
	\caption{Comparison between the GT optimization results, against high-fidelity outcomes for the GT-SOFC propulsion system with and without the battery, for the cases with and without heat exchangers by-pass valves. The black line indicates the perfect alignment between the optimization results and the high-fidelity outcomes, while the grey shaded area indicates $\pm$ 10.5$\%$ error region.}
	\label{fig:GT_validation_1}
\end{figure}
\begin{figure}[htbp]
	\centering
	\includegraphics[width=\linewidth]{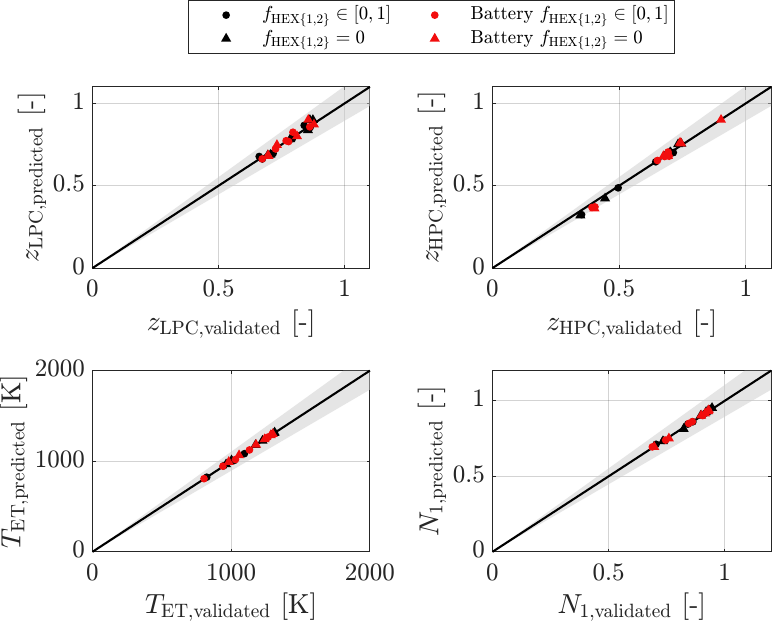}
	\caption{Comparison between the GT optimization results, against high-fidelity outcomes for the GT-SOFC propulsion system with and without the battery, for the cases with and without heat exchangers by-pass valves. The black line indicates the perfect alignment between the optimization results and the high-fidelity outcomes, while the grey shaded area indicates $\pm$ 10.5$\%$ error region.}
	\label{fig:GT_validation_2}
\end{figure}
\begin{figure}[htbp]
	\centering
	\includegraphics[width=\linewidth]{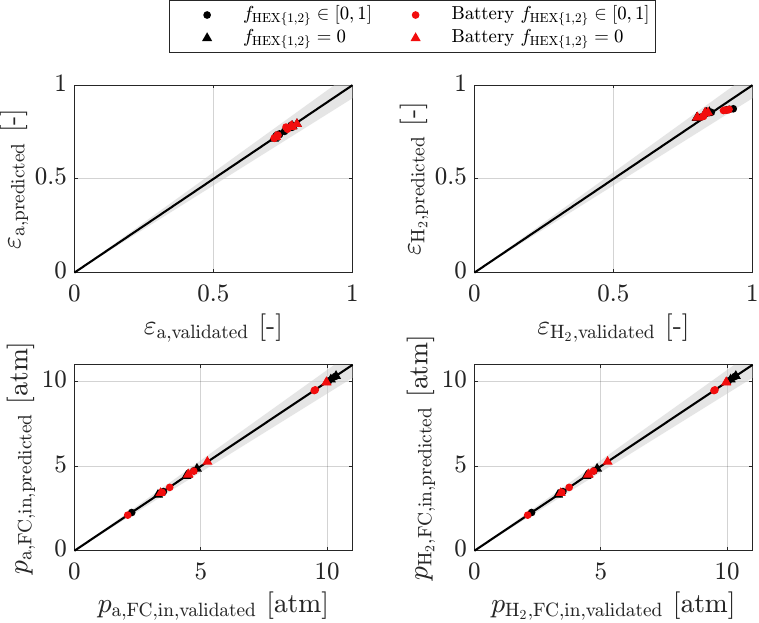}
	\caption{Comparison between the optimized values and the high-fidelity results for the air and fuel heat exchanger effectivenesses, $\varepsilon_\mathrm{a}$ and $\varepsilon_\mathrm{H_2}$, and the SOFC cathode and anode inlet pressures, $p_\mathrm{a,FC,in}$ and $p_\mathrm{H_2,FC,in}$, for the GT-SOFC propulsion system with and without the battery, for the cases with and without heat exchangers by-pass valves. The black line indicates the perfect alignment between the optimization results and the high-fidelity outcomes, while the grey shaded area indicates $\pm$ 7$\%$ error region.}
	\label{fig:HEX_validation}
\end{figure}
\begin{figure}[t]
	\includegraphics[width=\linewidth]{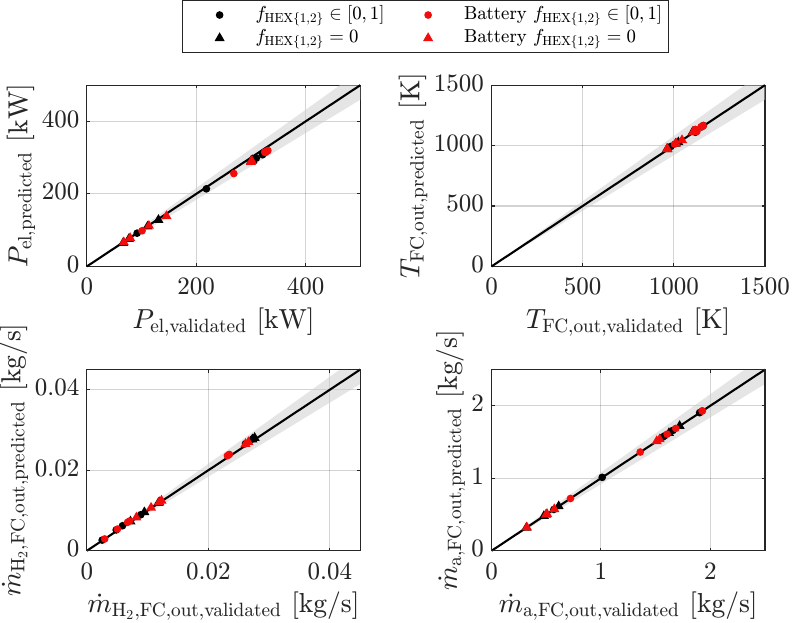}
	\caption{Comparison of SOFC optimization results, against high-fidelity outcomes for the GT-SOFC propulsion system with and without the battery, for the cases with and without heat exchangers by-pass valves. The black line indicates the perfect alignment between the optimization results and the high-fidelity outcomes, while the grey shaded area $\pm$ 8$\%$ error region.}
	\label{fig:SOFC_validation}
\end{figure}
\begin{figure*}[!htbp]
	\centering
	\begin{subfigure}[t]{0.49\textwidth}
		\centering
		\includegraphics[width=\linewidth]{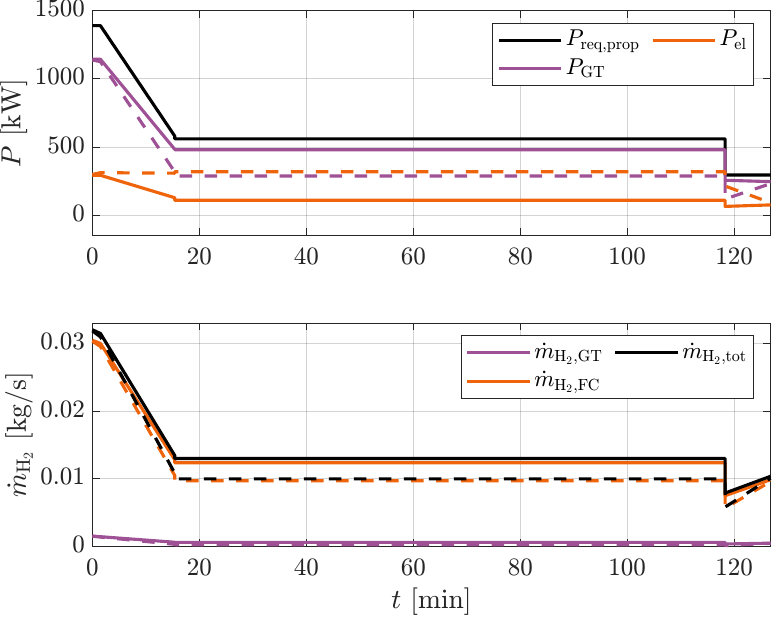}
		\caption{Comparison of GT and SOFC power and fuel flows for the GT-SOFC propulsion system with and without the heat exchangers by-pass valves.}
		\label{fig:sub1_GT_SOFC}
	\end{subfigure}
	\hfill
	\begin{subfigure}[t]{0.49\textwidth}
		\centering
		\includegraphics[width=\linewidth]{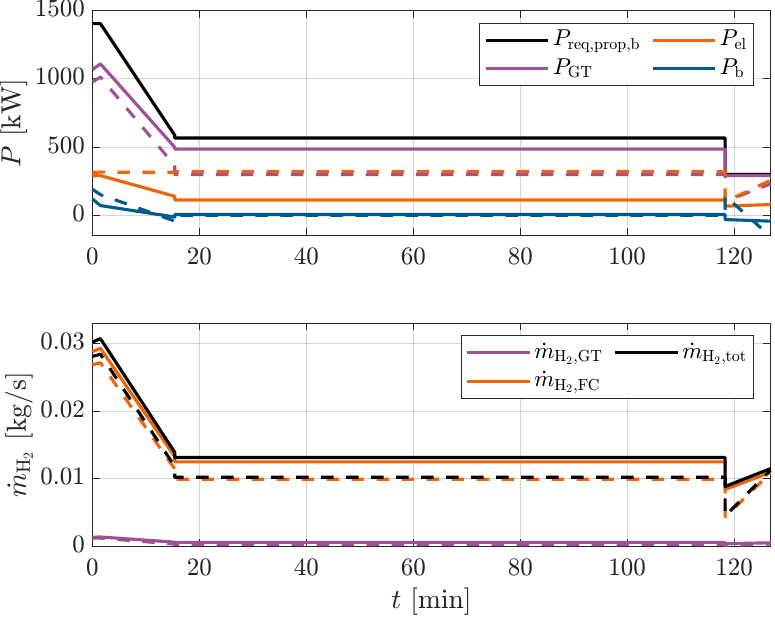}
		\caption{Comparison of GT, SOFC, and battery power and fuel flows for the GT-SOFC propulsion system integrating a battery with and without the heat exchangers by-pass valves.}
		\label{fig:sub2_GT_SOFC_BAT}
	\end{subfigure}
	\vspace{2mm}
	{\centering
		\parbox{1\textwidth}{\caption{Comparison of power and fuel flows for: (a) the GT-SOFC propulsion system with and without the heat exchangers by-pass valves and (b) GT-SOFC propulsion system integrating a battery with and without the heat exchangers by-pass valves. Continuous lines refer to the system without by-pass valves, whereas dashed lines refer to the system with the the by-pass valves.}
			\label{fig:GT_SOFC_P_mf}}
		\par}
\end{figure*}

\subsection{Validation} 
In order to validate our models, we feed the optimization results obtained from Problem~\ref{prob:Prob1} and Problem~\ref{prob:Prob2} in the high-fidelity models of GT, HEX1 and HEX2, and SOFC. Fig.~\ref{fig:GT_validation_1} and Fig.~\ref{fig:GT_validation_2} show that the GT results exhibit a relative error below 8$\%$, except for the high-pressure compressor surge margin indicator $z_\mathrm{HPC}$ during the FID 1500 ft condition for the baseline configuration integrating the battery, for which the relative error reaches approximately 10.5$\%$. Similarly,  Fig.~\ref{fig:HEX_validation} shows the effectiveness and outlet pressure results for the HEX1 and HEX2. In this case, the relative error remains below 7$\%$. Fig.~\ref{fig:SOFC_validation} shows that for all the four propulsion configurations, we obtain a relative error below 8$\%$ comparing the SOFC results against the high-fidelity outcomes. Despite covering a wide operating range and depending on several independent variables, the proposed models achieve good agreement with the high-fidelity simulation results, with a relative error of all variables below 10.5$\%$. 
\subsection{Numerical results}
\begin{figure}[htbp]
	\centering
	\centering
	\includegraphics[width=\linewidth]{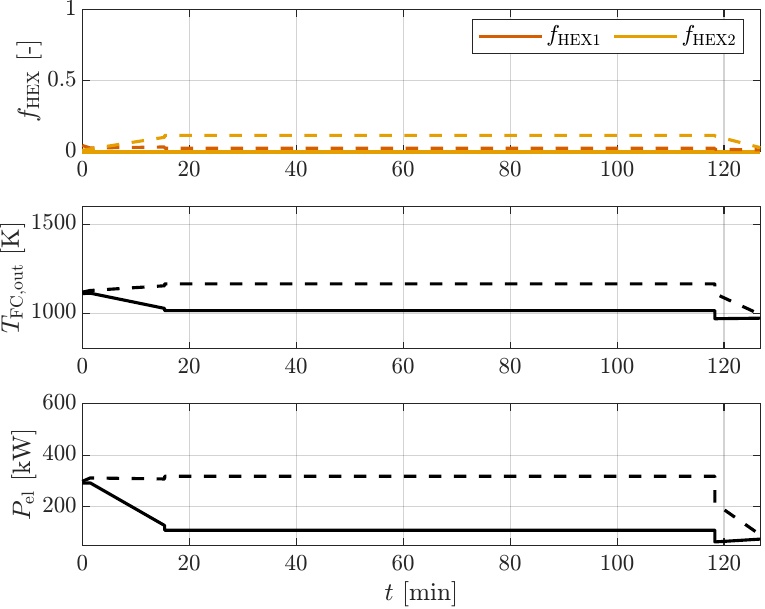}
	\caption{Heat exchangers by-pass valves opening and comparison of SOFC off-gases outlet temperature $T_\mathrm{FC,out}$ and electrical power $P_\mathrm{el}$ for the GT-SOFC propulsion system with and without heat exchangers by-pass valves. Dashed lines correspond to the configuration with bypass valves, while solid lines indicate the configuration without bypass valves.}
	\label{fig:T_fc_out_Pcell}
\end{figure}

\begin{figure}[htbp]
	\centering
	\centering
	\includegraphics[width=\linewidth]{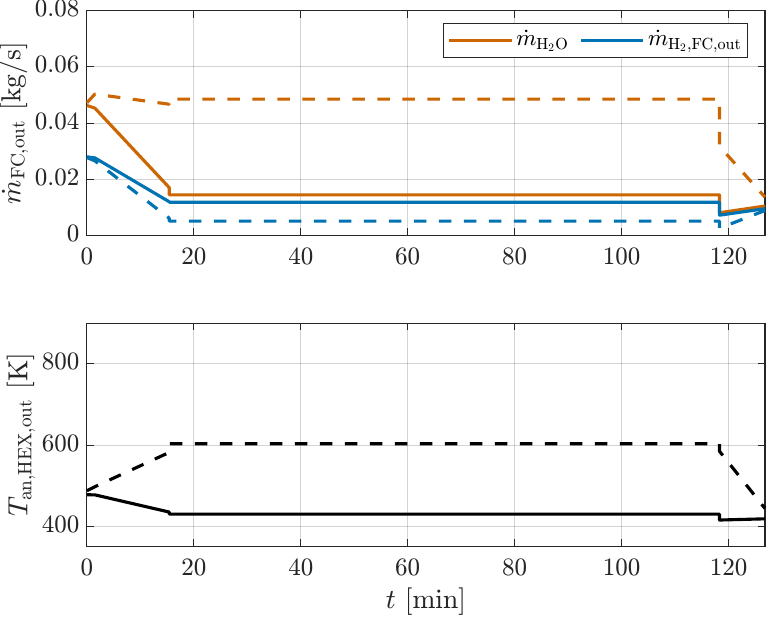}
	\caption{Comparison of SOFC off-gases hydrogen and steam content, and H$_2$ HEX hot side outlet temperature $T_\mathrm{an,HEX,out}$ for the GT-SOFC propulsion system with and without heat exchangers by-pass valves. Dashed lines correspond to the configuration with bypass valves, while solid lines indicate the configuration without bypass valves.}
	\label{fig:m_T_anode}
\end{figure}

\begin{figure}[htbp]
	\centering
	\centering
	\includegraphics[width=\linewidth]{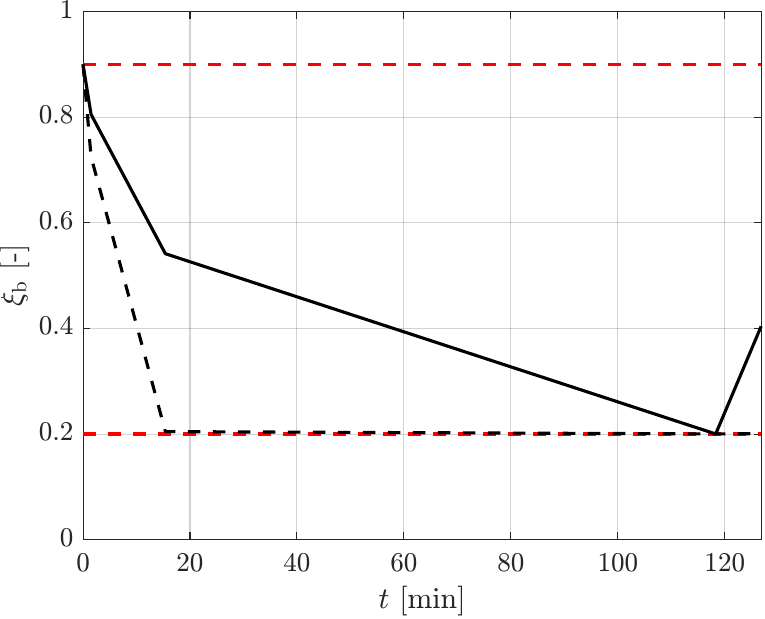}
	\caption{Comparison of battery state-of-charge $\xi_\mathrm{b}$ for the GT-SOFC-Battery propulsion system with and without HEX by-pass valves. The black dashed line corresponds to the configuration with bypass valves, while the solid line indicates the configuration without bypass valves. The red dashed lines denote the minimum and maximum state-of-charge levels.}
	\label{fig:SoC}
\end{figure}

\begin{figure}[H]
	\centering
	\includegraphics[width=\linewidth]{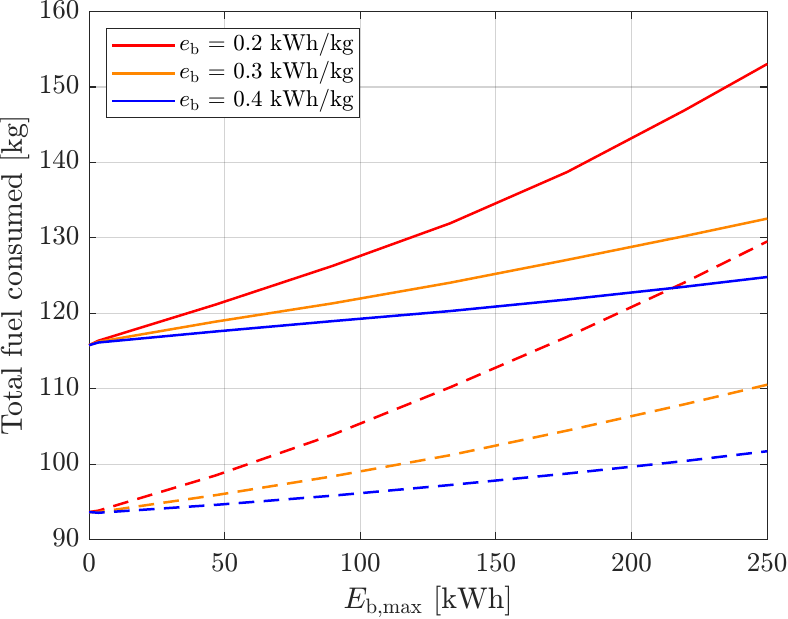}
	\caption{Comparison of optimal total fuel consumption for different battery sizes and energy densities. Dashed lines correspond to the configuration with bypass valves, while solid lines indicate the configuration without bypass valves.}
	\label{fig:Bat_eb_mf}
\end{figure}

After solving Problem~\ref{prob:Prob1}  for the cases with and without the heat exchangers by-pass valves, we obtain a total fuel consumption of \SI{115.78}{\kilogram} and \SI{93.66}{\kilogram}, respectively, leading to an approximate 19.11$\%$ fuel reduction over the same flight mission in favor of the configuration integrating two by-pass valves, one per heat exchanger, as shown in the middle subplot of Fig.~\ref{fig:sub1_GT_SOFC}. This improvement is explained by the enhanced temperature control enabled by the heat exchangers by-pass valves. We observe that the SOFC inlet temperature, which is limited to a maximum value for safety reasons, represents a limit for the SOFC off-gases outlet temperature. When by-pass valves are available, part of the flow can by-pass the corresponding HEX, as shown in the upper subplot of Fig.~\ref{fig:T_fc_out_Pcell}, allowing the SOFC to operate at higher temperatures while still respecting the inlet temperature limit. Operating at higher temperatures reduces the electrochemical losses and increases the SOFC power for the same operating voltage. In addition, as shown in Fig.~\ref{fig:m_T_anode}, the overall off-gases mass flow composed of steam and residual hydrogen increases, together with the outlet H$_2$ HEX temperature, leading to an higher enthalpy flow injected within the combustion chamber of the GT and to an enhanced system efficiency.
Similarly, after solving Problem~\ref{prob:Prob2} for the cases with and without by-pass valves, we obtain an approximate 19.56$\%$ total fuel reduction over the same flight mission in favor of the configuration including heat exchangers by-pass valves, as shown in the bottom subplot of Fig.~\ref{fig:sub2_GT_SOFC_BAT} and as reported in Table~\ref{tab:Results}. For the same reason explained above, the introduction of two by-pass valves allows to have a better temperature control, exploiting the SOFC more and having an enhanced heat recuperation.
Moreover, we observe that although the required propulsive power $P_\mathrm{req,prop}$ increases due to the additional weight of the battery, the total fuel consumption increases only by about 1.03$\%$ compared to the system not integrating the battery for the case without by-pass valves, while it increases by about 0.48$\%$ for the case with the valves. This outcome is justified by the fact that, intuitively, the battery discharges, contributing to meet the required power demand,  as shown on Fig.~\ref{fig:SoC}, therefore reducing the power that must be delivered by the GT and SOFC, for both the cases with and without by-pass valves. Furthermore, we observe that the by-pass valves not only modify the thermodynamic couplings, but also increase battery discharge during the initial flight phases, which are the the most demanding in terms of power. At the same time, they reduce the excess power generated by the SOFC and the GT during descent, thereby limiting battery recharge. However, the battery energy density has a relevant impact on the total fuel consumption, due to the additional weight given by the battery. Fig.~\ref{fig:Bat_eb_mf} shows how the optimal total fuel consumption changes as function of different battery capacities $E_\mathrm{b,max}$ and energy densities $e_\mathrm{b}$, accounting for the additional corresponding weight, for both propulsion systems (with and without the by-pass valves). We observe that integrating a small battery with current state-of-the-art technology ($e_\mathrm{b}$ ranging between 0.15 and \SI{0.25}{\kilo\watt\hour/\kilogram} \cite{PattanayakEtAl2025}) yields a slight increase in total fuel consumption compared to the system without a battery, for both the cases with and without by-pass valves. Conversely, the additional mass of larger batteries remains a limiting factor for their effective application, leading to a significant fuel consumption increase.
\begin{table}[htbp] 
	\centering 
	\caption{Comparison of the total fuel consumption of the baseline GT-SOFC propulsion system with that of the the same system integrating: (i) a battery
		only, (ii) heat exchangers by-pass valves only, and
		(iii) both a battery and heat exchangers by-pass
		valves.} 
	\label{tab:Results} 
\resizebox{\columnwidth}{!}{	\begin{tabular}{lccc} 
		\hline
		\centering
		\, & \multicolumn{2}{c}{\textbf{Fuel consumption [kg]}} 
		\\
		& $f_{\mathrm{HEX}\{1,2\}} = 0$ & $f_{\mathrm{HEX}\{1,2\}} \in [0,1]$ 
		& $\Delta_{f_{\mathrm{HEX}\{1,2\}}}$ \\
		\hline
		GT-SOFC & 115.78 & 93.66 & - 19.11$\%$  \\
		GT-SOFC-Battery & 116.99 & 94.11 & - 19.56$\%$  \\
		$\Delta_\mathrm{battery}$ & + 1.03$\%$ &+ 0.48$\%$ \\
		\hline
	\end{tabular}}
\end{table}
\begin{table}[htbp] 
	\centering 
	\caption{Comparison of the optimal total fuel consumption of the baseline GT-SOFC propulsion system with by-pass valves and the same system operated at constant degree of hybridization.} 
	\label{tab:Results_const_SOFC_power} 
	\resizebox{\columnwidth}{!}{	\begin{tabular}{lccc} 
			\hline
			\centering
			\, & \multicolumn{2}{c}{\textbf{Fuel consumption [kg]}} 
			\\
			& Minimum-fuel & Constant DoH 
			& $\Delta_{{\mathrm{minimum-fuel}}}$ \\
			\hline
			 GT-SOFC & 93.66 & 112.76 & - 16.94$\%$  \\
			\hline
	\end{tabular}}
\end{table}%
The thermodynamic integration, which is even more pronounced using the two heat exchangers by-pass valves, leads to a significant reduction in fuel consumption compared to the reference GT-powered propulsion system operated over the same mission, whose optimal fuel consumption is approximately \SI{129.55}{\kilogram}. Furthermore, in Table \ref{tab:Results_const_SOFC_power} we compare the minimum-fuel operation of the baseline architecture with by-pass valves to a constant degree of hybridization strategy, a simple and practically implementable benchmark strategy, where the ratio between the power produced by the SOFC and the total requested propulsive power is maintained constant throughout the mission. The constant degree of hybridization is set to the maximum feasible value that satisfies all constraints throughout the mission. The results show that the minimum-fuel strategy achieves a 16.94$\%$ reduction in total fuel consumption compared to the constant DoH case, highlighting the impact of the operating strategy on fuel consumption. However, the constant DoH strategy leads to a more stable thermal behaviour of the stack, with reduced variations of $\Delta T$ across it, thereby limiting thermal stresses.
\FloatBarrier

%% file: Sections/conclusions.tex
\section{Conclusions}
In this paper, we proposed reduced-order surrogate models for components to compute and compare minimum-fuel optimal operating strategies for hydrogen-based hybrid-electric aerospace propulsion systems under steady-state conditions. In particular, we analyzed the reference configuration and compared it with three modified configurations integrating: (i) a battery, (ii) two heat exchangers by-pass valves, and (iii) both the battery and by-pass valves.  Our results show that employing the two heat exchangers by-pass valves leads to a significant fuel reduction---19.11$\%$ and 19.56$\%$ for the configurations with and without the battery, respectively---as it allows to exploit the solid-oxide fuel cell more, which is more efficient than the gas turbine. Moreover, the comparison with a constant degree of hybridization strategy highlights a trade-off between fuel efficiency (reduction of 16.94$\%$ in fuel consumption) and a more conservative SOFC operating regime, associated with more stable thermal conditions across the stack. Additionally, although the use of a small advanced battery leads to a slight fuel consumption increase under steady-state conditions compared to the corresponding reference configuration, it is expected to have a significant advantage during transient operations, which are not captured in the present work.
Overall, the proposed models and framework can aid design experts in selecting promising configurations and design points, such that the propulsion system operates efficiently throughout the entire flight mission, while striking a balance between accuracy and computational cost.

This takes us to the future research in this field: First, the proposed operating strategies could be integrated into real-time control schemes.	Second, the GT and SOFC should also match during transient operations, as the dynamic behavior of the two components is very different.
Finally, the proposed framework can be extended to study and minimize NO$_x$ emissions.